\documentclass[a4paper,11pt]{article}
\usepackage{jcappub}
\pdfoutput=1
\usepackage[english]{babel}
\usepackage{amsmath,amssymb,amsfonts, bm,bbm,slashed, subdepth}
\usepackage{graphicx,subfigure}
\usepackage[colorlinks=true]{hyperref}
\usepackage{cleveref}
\usepackage{enumerate}
\usepackage{setspace}
\usepackage{booktabs, tabularx}
\usepackage{units}
\usepackage{slashed}
\usepackage[dvipsnames]{xcolor}
\usepackage{soul}
\usepackage{comment}
\usepackage{multirow}

\makeindex

\begin{document}

\title{
Light from darkness: history of a hot dark sector
}

\author{Rupert Coy,}
\emailAdd{rupertlcoy@gmail.com}

\author{Jean~Kimus}
\emailAdd{jean.kimus@ulb.be}

\author{and Michel H.G. Tytgat}
\emailAdd{michel.tytgat@ulb.be}

\affiliation{Service de Physique Th\'eorique, Universit\'e Libre de Bruxelles, Boulevard du Triomphe, CP225, 1050 Brussels, Belgium}

\abstract{
We study a scenario in which the expansion of the Early Universe is driven by a hot hidden sector (HS) with an initial temperature $T'$ that is significantly higher than that of the visible sector (VS), $T' \gg T$. The latter is assumed to be made of Standard Model (SM) particles and our main focus is on the possibility that dark matter (DM) is part of the dominant HS and that its abundance is set by secluded freeze-out.  In particular, we study the subsequent evolution and fate of the DM companion particle after freeze-out all the way toward reheating of the VS. To make this scenario more concrete, we work within dark QED, a framework in which the DM is a Dirac fermion and its companion, a massive dark photon; coupling between the SM and HS is through kinetic mixing. We provide a detailed and comprehensive numerical and analytical analysis of the different regimes of reheating of the VS. Extending and complementing the work of Coy {\em et al} on the``domain of thermal dark matter candidates" \cite{Coy:2021ann}, we use our results to explore the viable parameter space of both the DM matter particle and its companion, here the dark photon. We show that current and future fixed target experiments can probe scenarios along which the expansion was driven by relativistic DM photons, a scenario dubbed relativistic reheating. We also set new bounds on the maximal temperature ratio $T'/T$ and argue for an extension of the domain toward very large DM masses, $m_{\rm dm} \sim 10^{11}$ GeV. These are possible assuming that DM annihilation is bounded by unitarity and that reheating of the VS occurs just before big bang nucleosynthesis. We also discuss some possible implications for (and constraints on) baryogenesis, including simple leptogenesis mechanisms, and how they may set additional constraints on the domain of DM candidates.}

\maketitle

\section{Introduction}

Despite intensive experimental and theoretical efforts, the nature of dark matter (DM) remains elusive. In recent years, much focus has been put on the possibility that DM is a particle that belongs to a hidden sector (HS).
There is no universal definition of a HS, but its constituents must be Standard Model singlets that interact at most feebly with the Standard Model (SM), or visible sector (VS) . 
The HS may be very complex, leading to a rich and fascinating phenomenology, with signatures from colliders to astrophysics and cosmology \cite{Foot:1991bp,Strassler:2006im,Curtin:2017quu,Spergel:1999mh,Tulin:2017ara,Schwaller:2015tja}. 
From a cosmological perspective, it is reasonable to conceive that the HS and the VS were not in thermal equilibrium with each other during all stages of the Universe's evolution, if at all. Consequently, they could have evolved with quite different temperatures, $T' \neq T$. A popular category of scenarios assumes that the HS is initially depleted compared to the VS, itself being populated at the end of inflation through the reheating process \cite{Kolb:1990vq}. In such scenarios, the DM particles, and possibly their companions, could have been created through the freeze-in mechanism \cite{Dodelson:1993je,McDonald:2001vt,Hall:2009bx,Chu:2011be,Bernal:2017kxu}. This may have happened in many ways, in particular through the  Standard Model portals \cite{Patt:2006fw}, including gravity \cite{Kolb:2023ydq, Clery:2021bwz}. In such scenarios, depending on the interactions of the DM and its companions, the HS may have reached thermal equilibrium, but typically with $T' \ll T$.

In this work, instead, we study scenarios in which the HS was initially more populated than the VS.  Specifically, we will assume that the HS and VS were in thermal equilibrium but with $T ' \gg T$ \cite{Berlin:2016vnh,Heurtier:2019eou,Ertas:2021xeh,Garani:2021zrr,Coy:2021ann,Bringmann:2023iuz}. Of course, this must end before the time of big bang nucleosynthesis (BBN), but at early times, it is conceivable if the inflaton decayed dominantly into HS particles and much less in VS particles, a scenario called asymmetric reheating \cite{Berezhiani:1995am,Adshead:2016xxj,Hardy:2017wkr,Tenkanen:2016jic,Sandick:2021gew,Ireland:2022quc,Ghosh:2023mnh}. 

Our main motivation to consider a relatively hot HS at early times is to complement the work done in \cite{Coy:2021ann} in which the possible DM particle candidates that had been in thermal equilibrium in the Early Universe were delineated in the plane $m_{\rm dm}$ vs  the initial temperature ratio between the hidden and visible sectors, $\xi_i = (T'/T)_{\rm init.}$. This is called 'the domain' in the sequel. That work used very general constraints and arguments, like requiring that the annihilation cross section of the DM is bounded by unitarity \cite{Griest:1989wd}. It was also assumed that possible companions of the DM particles were subdominant and/or played only a secondary role. In the present work, our goal is to determine to which extent this is indeed the case and to study the possible implications, both for the properties of the companions and for the domain of DM candidates. Our main results in that regard are depicted in figure \ref{fig:newDomain} that shows the domain taking into account different possible outcomes for the DM companion. Perhaps the most striking feature is that the DM mass range extends to much larger values, $m_{\rm dm} \sim 10^{11}$ GeV than in \cite{Coy:2021ann}. This is possible if, along the process of heating of the VS, the companions release the largest possible amount of heat or entropy right before BBN, a rather extreme scenario. This new bound confirms already existing results \cite{Berlin:2016vnh,Heurtier:2019eou,Garani:2021zrr,Bernal:2023ura}, but with a distinct focus and a from a quite different perspective. 

Concretely, our setup is as follows. We will first assume that the Universe has two independent sectors, HS and VS, both in thermal equilibrium but characterised  by a constant temperature ratio initially, $\xi_i =  T'_i/T_i$. We will focus in particular on the case of $\xi_i \gg 1$ but, for continuity, consider also the opposite, $\xi_i \ll 1$. Second, we will assume that the DM abundance is  determined by thermal freeze-out within the HS. The case of relativistic DM freeze-out, similar to the freeze-out of SM neutrinos, defines what has been called the 'relativistic floor' in \cite{Hambye:2018dpi,Coy:2021ann}, see figure \ref{fig:newDomain}. Relativistic freeze-out occurs typically provided the DM companions are heavier than the DM particles. The bulk of the domain corresponds to DM particles whose abundance is set by annihilation into lighter HS particles, a scenario called secluded freeze-out \cite{Pospelov:2007mp}. Our main problem throughout will be to follow the subsequent evolution of the companion particles after DM freeze-out, making sure that the Universe is radiation dominated and made dominantly of SM particles by the time of BBN \cite{Hannestad:1995rs,Kawasaki:2000en}. 

Both to be concrete and for its own sake, we will explore this non-standard cosmological history within a very popular HS model, meaning 'dark QED' ~\cite{Ackerman:2008kmp,Feng:2009mn,Arvanitaki:2021qlj}. In that model,  the DM is made of dark electrons and positrons (DM, mass $m_{\rm dm}$) that interact through a dark photon which we assume is massive (DM companion, mass $m'$). One of the appealing features of dark QED is that DM stability is related to dark fermion number conservation \cite{Hambye:2010zb}. As usual, this HS and the SM are  coupled through  kinetic mixing (mixing parameter $\varepsilon$)~\cite{Holdom:1985ag,Patt:2006fw}. This setup is very simple but is representative of a large class of models for a HS with DM \cite{Chu:2011be,Essig:2015cda}. While it has been extensively studied both theoretically and experimentally, it has not yet been explored in the context we are considering. Moreover, given the interest in dark photons \cite{Cline:2024qzv}, we will specifically focus on the constraints or requirements that a viable hot HS imposes on the parameters of the dark photon, $\varepsilon$ and $m'$, as shown in figure \ref{fig:constraints}.

We aimed to draw general conclusions from our work, particularly regarding the domain of DM candidates, which has a cost. Specifically, we found it necessary to systematically study quite distinct regimes, some well-known and others not yet explored in the literature. On top of numerical results, we provide in all cases analytical solutions, which we then used to update the domain, as shown in figure \ref{fig:newDomain}. We also derived several intermediate results, some of which are given in the appendices. Our work is consequently a bit lengthy. A reader mostly interested in the main conclusions and phenomenological considerations may go directly to sections \ref{sec:constraints} and  \ref{sec:domain} for implications for the properties of dark photons and for the domain of thermal dark matter candidates. Finally, while we tried to be self-contained, it may be useful to read in parallel the results reported in \cite{Coy:2021ann}.

Most of our work revolves around the problem of reheating the VS from a relatively hot HS. After freeze-out, the dark photons (or dark matter companions) remain essentially relativistic and very abundant, but due to kinetic mixing, they eventually decay into SM particles, reheating the VS in the process. As we will show, the reheating process can occur in two distinct ways. First, it may happen when the dark photons are  relativistic. We will refer to this as `relativistic reheating'. To our knowledge, this possibility has not been studied in details, if at all, in the literature. The possibility of relativistic reheating depends on the properties of the companion particles. In particular, they should not be too heavy and too feebly coupled to the SM. As figure \ref{fig:constraints} shows, current and forthcoming fixed target experiments, notably SHiP, are able to probe such cosmological ingredients. Alternatively, reheating may take place when the dark photons become non-relativistic. We refer to such scenarios as `non-relativistic reheating'. This situation is, to a large extent, textbook material, starting with the seminal work of Scherrer and Turner. \cite{Scherrer:1984fd}. However, the conditions for non-relativistic reheating in our setup differ from those usually encountered in the literature. Consequently, we found it necessary to revisit the problem of entropy production to adapt it to our scenarios. We derive several approximate analytical expressions, some of which are new and which we then used to extend the domain, as depicted in figure \ref{fig:newDomain}. Along our analysis, we found that the process of heating of the VS is essentially controlled by a combination of decay and expansion rates and energy densities that  we call the 'heating parameter', $\kappa$, see eq.\eqref{eq:heatpar}. In particular, $\kappa \sim 1$ between the onset and the end of the reheating process, see figure \ref{fig:XiSchematic} for a schematic representation. 

 One key question regarding  the domain of thermal DM candidates is whether there is an upper bound on the ratio of temperatures between the HS and VS? In \cite{Coy:2021ann}, such bound $\xi_i$ has been set by assuming that all the HS particles were subdominant by the time of BBN, using constraints on $\Delta N_{\rm eff}$. In the present work, we set stronger and, arguably, more realistic bounds by requesting that a constant temperature ratio $\xi \approx \xi_i$ is a well-defined initial condition, see section \ref{sec:stability}. 

Besides the DM problem, a relatively hot HS may have other implications. For instance,  gravitational waves from a phase transition that occurred when the Universe is dominated by a HS have been considered in \cite{Fairbairn:2019xog,Ertas:2021xeh,Bringmann:2023iuz}. In this work, we briefly discuss some possible implications for baryogenesis and leptogenesis \cite{Mazumdar:2013gya,Davidson:2002qv}. 
On top of bounds set on the maximal temperature ratio if the baryon asymmetry is created before reheating of the VS, we discuss two aspects that are rather generic and that rest on the much faster expansion rate of a Universe dominated by a hot HS. First, we  consider the freeze-out of sphalerons around the electroweak phase transition, a feature relevant for electroweak baryogenesis \cite{Davidson:2002qv}. Second, we  generalize the Davidson-Ibarra bound \cite{Davidson:2000dw} on the mass of heavy neutrinos for leptogenesis mechanisms. In particular, we argue that viable leptogenesis requires that the temperature ratio be less than $\xi_i \sim 10^{2}$, see figure \ref{fig:newDomain}.

\bigskip
Our work is organized as follows.  We establish our groundwork in section \ref{sec:history}, in which we briefly recap the features of dark QED  and write down a set of simple Boltzmann equations that describe the evolution of the hidden and visible sectors, including entropy production. 
In section \ref{sec:BoltzSolns}, we give several analytical solutions,  studying the evolution of the temperature ratio, $\xi \gg 1$ towards standard cosmological evolution. 
We identify several possible scenarios that depend critically on whether reheating occurs in a radiation (relativistic reheating) or matter dominated (non-relativistic reheating) era.  
In section \ref{sec:constraints}, we delineate the parameter space of the dark photon in the plane $\varepsilon$ vs $m'$, the dark photon mass, see figure \ref{fig:constraints}. 
In section \ref{sec:domain}, we discuss the implications of reheating on DM and present an updated domain of thermal DM candidates, going beyond the specific dark QED model, see figure \ref{fig:newDomain}. That section also covers the case of a cold HS ($\xi_i \ll 1$). 
In section \ref{sec:baryo}, we  discuss some possible implications of a hot HS, in particular for the baryon asymmetry of the Universe and for two basic mechanisms, leptogenesis and the freeze-out of sphaleron process for electroweak baryogenesis and put extra bounds on $\xi_i$.  
We finally draw our conclusions. The problem of entropy production during energy transfer from the HS to VS  is revisited in details in appendix \ref{app:entropyProduction}.  In appendix \ref{app:therm}, we give a summary of Maxwell-Boltzmann relations, which we use throughout, including for relativistic particles. Other appendices cover additional technical details. 
{Throughout this work, we use primes to denote quantities in the hidden sector (e.g., the entropy density $s^\prime$), and unprimed symbols for those in the visible sector. We use the subscript 't' to refer to the sum of these quantities. For example the total entropy density is written as $s_t = s^\prime + s$.}

\section{Basic HS ingredients}
\label{sec:history}

\subsection{Dark QED with kinetic mixing}

We consider a HS that consists of dark QED. 
DM is made of a  Dirac fermion $\chi$ (and its anti-particle  - we do not consider the possibility of asymmetric dark matter) and the companion is a massive dark photon ($\gamma'$) ~\cite{Ackerman:2008kmp,Feng:2009mn}. For the purpose of efficient DM secluded freeze-out, we assume that the dark photon is  lighter than the DM, $m' \lesssim m_{\rm dm}$. Finally, the DM and dark photons interact with the SM through kinetic mixing \cite{Holdom:1985ag}. The Lagrangian is  
\begin{eqnarray}
\label{eq:lag}
{\cal L} &\supset&
\overline{\chi} (i \slashed{D} - m_{\rm dm}) \chi - \frac{1}{4} F'_{\mu \nu} F'^{\mu \nu} \nonumber\\
&+& {1\over 2} m'^2 A'_\mu A'^\mu- \frac{\varepsilon}{2} B_{\mu \nu} F'^{\mu \nu} \, ,
\end{eqnarray}
where $B_{\mu \nu}$ is the SM hypercharge field strength. The covariant derivative involves a  coupling $e'$, hence a HS fine structure constant, $\alpha' = e'^2/4\pi$. 
The dark photons can acquire a mass via the Stueckelberg or the Brout-Englert-Higgs mechanism. We assume  that the details of $U(1)'$ symmetry breaking play only a subsidiary role for the fate of the dark photons. 
The mixing term $B_{\mu \nu} F'^{\mu \nu}$ requires a redefinition of the gauge boson fields in order to write down canonical gauge kinetic terms. There are some subtleties regarding the limit of a massless dark photon, but they play little role in the body our work, see e.g. \cite{An:2013yfc,Hambye:2019dwd}.

\subsection{History of a hot hidden sector in brief}
\label{sec:problem}

There are several ways to produce DM through kinetic mixing \cite{Chu:2011be,Hambye:2019dwd,Arvanitaki:2021qlj}.  
In these references, it is assumed that the HS is either subdominant or is in thermal equilibrium with the VS. Here, instead, we suppose that the Universe is dominated by the HS at early times.  We parameterize this by assuming thermal equilibrium in both sectors, characterized by temperatures $T'$ and $T$. Thus we assume  $\xi_i = T_i'/T_i \gtrsim 1$ initially.  While dark QED has  fewer degrees of freedom than the SM, a large initial $\xi$ generically implies {that the HS energy density $\rho'$ dominates over the VS one,}  $\rho' \gg \rho$. For $g'_\ast = 3$ and $g_\ast = {\cal O}(100)$, $\rho' \gtrsim \rho$ provided $\xi \gtrsim (g_\ast/g_\ast')^{1/3} \sim 2.4$. 

\begin{figure*}[t!]
    \centering
    \includegraphics[width=15cm]{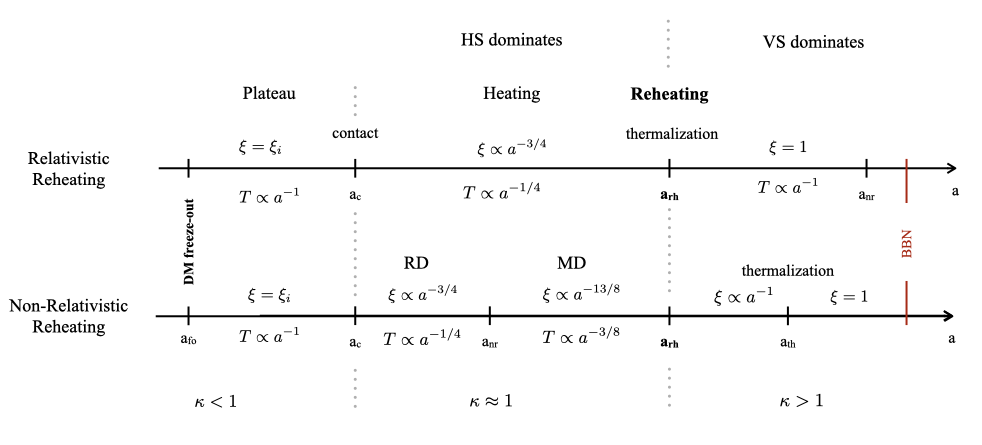}
    \caption{
Characteristic evolution of the temperature ratio, $\xi = T'/T \gg 1$, the VS temperature $T$, and `heating parameter', $\kappa = (\rho'/3 \rho) \times \langle \Gamma'\rangle/ H$, as functions of the scale factor $a$, starting from a hot HS with $\xi_i \gg 1$ after the moment of DM freeze-out, $a_{\rm fo}$.  A well-defined scenario requires that the HS is subdominant by the time of BBN. This amounts to requiring that $m' \gtrsim $ MeV. Also, we require that  $\kappa \ll 1$ at early times,  $\xi \approx \xi_i$ to which we refer as the plateau. Energy transfer, and so 'heating' of the VS, effectively starts at the moment that we call `contact', when $\kappa$ reaches 1. The heating parameter remains ${\cal O}(1)$ throughout heating and starts to increase after heating is over and the VS is reheated. Relativistic reheating corresponds to parameters such that the dark photons are still relativistic at reheating, thermalize with the VS, with $T' = T$, and remain in thermal equilibrium afterward. Non-relativistic reheating corresponds to scenarios in which the dark photons become NR before the reheating. The dark photons remain out-of-equilibrium, but may thermalize if their abundance reaches the equilibrium one $n'_{\rm eq}(T)$.}
    \label{fig:XiSchematic}
\end{figure*}

This section is dedicated to a brief introduction of the stages from $\xi_i \gg 1$ to reheating  before the time of BBN. The details are provided in the subsequent sections but the different stages are shown schematically in figure \ref{fig:XiSchematic}. The initial condition $\xi_i \neq 1$ is possible only provided the HS and VS are effectively decoupled. As we shall see in section \ref{sec:constraints}, this requirement puts an upper bound on the kinetic mixing parameter, $\varepsilon \lesssim 10^{-6}$, see figure \ref{fig:constraints}.
The corollary of a small $\varepsilon$ is that the dark photons are long-lived. Since the HS is initially much more populated than the VS, we will impose that the dark photons are not too abundant by the time of BBN. This will put a lower bound on $\varepsilon$, also depicted in figure \ref{fig:constraints}.

We will require that the DM abundance is set by secluded annihilation of DM into dark photons, $\chi \bar \chi \leftrightarrow \gamma'\gamma'$ \cite{Pospelov:2007mp}.
If expansion is dominated by the HS then $H \sim g_\ast'^{1/2} T'^2/m_{\rm pl}$. 
Assuming instantaneous freeze-out, the DM relic abundance is then of the order of
\begin{equation}
Y_{\rm dm}' = {n_{\rm  dm}\over s'} \sim {x'_\text{fo}\over g_*'^{1/2} m_{\rm dm} m_{\rm Pl} \langle \sigma v \rangle}
\label{eq:DMabundance}
\end{equation}
where $n_{\rm dm} =  n_\chi + n_{\bar \chi}$ {and $x'_{\rm fo} = m_{\rm dm}/T'_{\rm fo}$.} In this expression, $T'_{\rm fo}$ is the freeze-out temperature of the HS and $\langle \sigma v\rangle$ is the thermally averaged DM annihilation cross section into dark photon \cite{Kolb:1990vq}, see appendix \ref{app:cross_sec}. The primes on $Y'_{\rm dm}$ is to emphasize the fact that this expression gives the DM abundance as long as the entropy, and thus the expansion, is dominated by the HS. However, due to reheating of the VS, an irreversible process, $Y'$ is in general not the final DM abundance. 

After DM freeze-out, the DM abundance is in general Boltzmann suppressed and so $n_{\rm dm} \ll n_{\gamma'}$.  The Universe is thus essentially composed of dark photons, $n'\approx n_{\gamma'}$. A self-interacting companion particle, like a non-abelian gauge boson \cite{Garani:2021zrr} or self-interacting scalar \cite{Tenkanen:2016jic}, could remain in thermal equilibrium after DM freeze-out, but the dark photons can only interact with DM particles, which are rare after freeze-out. The dark photons are therefore kinetically decoupled and are free streaming.
As long as the dark photons remain relativistic and dominate the Universe, we may expect that their distribution is close to the one at equilibrium. While we will track deviation from thermal equilibrium in our work, we will see that it is useful to adopt the temperature $T'$ as a proxy to characterize the dark photon abundance as long as they are relativistic, $n' \propto g'_\ast T'^3$. Initially, the temperature of the dark photons evolves as $T'\propto a^{-1}$. However, through kinetic mixing, they will decay and heat the VS. 

The dominant process for heating is dark photon decay into pairs of SM particles, with a rate $\langle \Gamma' \rangle \sim m' \Gamma'/T'$ where the factor $m'/T' \ll 1$ is caused by time dilation. Thus decays are initially rare but become more and more relevant as time goes by. As we will emphasize, the efficiency of energy transfer from the HS to the VS is controlled both by the decay and expansion rates but also by the ratio of HS and VS energy densities, through $\kappa \sim (\langle\Gamma'\rangle/H)(\rho'/\rho)$. We will call ``heating" the early phase of energy transfer between the HS and the VS and call $\kappa$  the ``heating parameter" (see section \ref{sec:Heating_param}, see also \cite{Fairbairn:2019xog}). Briefly, we will show that the HS and VS are effectively decoupled  provided $\kappa \ll 1$ at early times and that heating effectively starts when $\kappa$ reaches and, interestingly, remains ${\cal O}(1)$ until heating is over, see figure \ref{fig:XiSchematic}. 

From the onset of heating, there are two possibilities. First, the temperature ratio may reach $\xi = T'/T = 1$ while the dark photons are still relativistic. Adapting a popular nomenclature, we will call this scenario ``relativistic heating" and use ``reheating" to refer to the end of the heating process. Provided $g'_\ast \ll g_{\ast}$, as is the case for dark QED, at reheating the bulk of the energy lies within the VS, $\rho\gg \rho'$. Reheating itself occurs when the heating parameter departs from $\sim 1$ and starts increasing again, $\kappa \gtrsim 1$. Since $\rho'/\rho$ is constant, this condition corresponds to the standard requirement $\langle \Gamma'\rangle\gtrsim  H$, see figure \ref{fig:XiSchematic}. In this scenario of relativistic heating, the dark photons thermalize with the VS. We will call ``thermalization" such moment when both sectors reach thermal equilibrium with others. In the case of relativistic heating, reheating and thermalization coincide, $T_{\rm rh} = T_{\rm th}$. Since $\langle \Gamma' \rangle > H$ always after reheating, the dark photons remain in thermal equilibrium with the SM particles. Eventually, they become non-relativistic with an abundance that is Boltzmann suppressed. So, they are and remain subdominant after reheating.

Alternatively, the dark photons may become non-relativistic while the expansion is still dominated by the HS, $\xi \approx m'/T \gtrsim 1$. In that case, the Universe becomes matter dominated before reheating. The VS keeps being heated by rare dark photon decays until $\Gamma' \sim H$. This stops when the bulk of the dark photons decay, corresponding to $\Gamma' \sim H$ at which point $\rho > \rho'$. This scenario is standard, as it corresponds to the familiar heating through the decay of a massive particle \cite{Scherrer:1984fd}. We refer to this scenario as ``non-relativistic heating", see section \ref{sec:nonrel} and figure \ref{fig:XiSchematic}. After non-relativistic heating, the abundance of dark photons is exponentially suppressed. The dark photons can thermalize with the VS provided their abundance crosses their equilibrium abundance at temperature $T$, $n' \sim \exp(-t_{\rm th}/\tau')\sim n'_{\rm eq}(T_{\rm th})$, see \cite{Harvey:1981yk} and appendix \ref{app:FI}. Thus, in the case of non-relativistic dark photons, reheating and thermalization are separate events and $T_{\rm th} \ll T_{\rm rh}$. This is just a matter of principle and the abundance of dark photons is exponentially small after reheating. 

In either cases of relativistic or non-relativistic heating, reheating must occur before BBN with a Universe that is radiation dominated \cite{Kawasaki:2000en,Hannestad:2004px} and mostly composed of SM particles \cite{Dodelson:1992km,Hannestad:1995rs,Mangano:2005cc}. Reheating before BBN requires that the dark photons have either thermalized with the VS (relativistic heating) or decayed (non-relativistic heating) before BBN. If this is satisfied, the second condition, which can be expressed in terms of the effective number of neutrino degrees of freedom at BBN, $\Delta N_{\rm eff}$, requires that the dark photons are non-relativistic by the time of BBN, $m' \gtrsim$ MeV, cf Fig \ref{fig:constraints}. If both conditions are satisfied, then the early Universe may have been dominated by a hot HS, $\rho' \gg \rho$. This does not mean that the process of reheating of the VS from a hot HS is without consequences. First, most of the entropy of the Universe lies initially within the HS and must be transferred to the VS. Second, entropy may be created through the irreversible process of heating. 
While entropy production may lead to dilution of the DM abundance, eq.\eqref{eq:DMabundance}, both effects are relevant if there are relics within the VS that are generated before reheating, like a baryon asymmetry, topological defects, other forms of DM, primordial gravitational waves, etc. Given the relevance of entropy production and transfer, we provide for completeness several expressions for entropy production both in the following section and, from a different perspective, in the appendix \ref{app:entropyProduction}. As we will show, baring issues with the baryon asymmetry of the Universe, expansion by a hot HS may allow thermal DM particles to be as heavy as $\sim 10^{11}$ GeV, much heavier than standard bounds, see figure \ref{fig:newDomain} (see also \cite{Berlin:2016gtr,Heurtier:2019eou,Bernal:2023ura}).

\subsection{Boltzmann equations}

In this section, we introduce the set of Boltzmann equations (BE) that we use to follow the evolution of the coupled hidden and visible sectors.
We begin right after DM freeze-out. {The abundance of DM particles at that moment is given by $Y_{\rm dm} \approx n_{\rm dm}/s_{{\rm t}}  \ll 1${, with $s_{\rm t} = s' + s$ the total entropy density at that same moment, where  $s_t \approx s'$ if $\xi \gg 1$}. As the DM particles are both decoupled and subdominant}, we only consider Boltzmann equations that involve  the dark photons and the SM degrees of freedom. Since the mixing parameter must be small to ensure that the sectors are decoupled initially, the most important processes are those that occur to lowest order in $\varepsilon$. These are the dark photon decay and inverse decay processes, $\gamma' \leftrightarrow f \overline{f}$, with rate $\Gamma' \propto \varepsilon^2$ (see appendix \ref{app:cross_sec} for detailed expressions). 
Scatterings and annihilation processes, meanwhile, scale as $\varepsilon^4$, and we have explicitly checked that they are indeed always subdominant.  
Also, processes like $\gamma' + f  \leftrightarrow \gamma + f$, while $\propto \varepsilon^2$, are  subdominant initially because $n \ll n'$. While they may play a subsidiary role  when $\xi = T'/T$ approaches $1$ and $n \sim n'$, thus possibly accelerating the process of reheating. For simplicity, we neglect such processes in our work. 

We  consider Boltzmann equations that are obtained by taking moments of the kinetic equations for the particle distributions, $f^{(\prime)}(\vec p,t)$, hence we use a fluid approximation.
Since we are concerned with evolution before BBN, we assume that the visible sector is in thermal equilibrium and radiation-dominated (RD). It it thus characterized by its effective number of degrees of freedom $g_\ast(T)$ and its temperature $T$. 
Somewhat paradoxically, the treatment of the HS is complicated by the fact that, after DM freeze-out and before reheating,  the dark photons are free streaming. 
At the end, in this work, we make two simplifying assumptions. First, we are assuming that the HS distribution remains close to equilibrium and so can be characterized by the temperature $T'$ and a chemical potential $\mu'$ to parameterize departure from thermal equilibrium. Second, we will use Maxwell-Boltzmann (MB) statistics, {\em including} for relativistic particles. We do so for the following reasons. First, while quantum statistics effects are in principle always relevant for relativistic particles, the numerical errors made using MB instead of Fermi-Dirac or Bose-Einstein statistics are in practice small, ${\cal O}(10\%)$ for equilibrium quantities. Second, admitting such numerical errors is more than compensated for by the ease with which MB equilibrium thermodynamical  quantities can be expressed over the whole temperature range.  For instance, the number density of dark photons is of the form $n' = n'_{\rm eq}(T') e^{\mu'/T'}$ with $n'_{\rm eq}(T')$ given in terms of simple Bessel functions  \cite{Gondolo:1990dk}, cf appendix \ref{app:therm}. Given all these approximations, in full generality, the coupled HS and VS can be characterized by three quantities, $T'$, $\mu'$ and $T$. Accordingly, we could have written Boltzmann equations directly for these quantities, as in \cite{Bringmann:2020mgx}. Here, instead, we consider equations for  $n'$, $\rho'$ and $\rho$. We then use MB equilibrium statistics to extract other quantities, like $T'$ and $T$ or entropy densities. 

We consider the following set of Boltzmann equations
\begin{eqnarray}
\frac{d\rho_{\rm t}}{dt}\!\!\! &+& \!3(1 + w_{\rm t}) H \rho_{\rm t} = 0
    \label{eq:continuity}\\
    \frac{dn'}{dt}\!\! &+&\!\! 3 H n' =  -\Gamma'\!\left(   \frac{K_1(x')}{K_2(x')} n' -\frac{K_1(x)}{K_2(x)} n'_{\rm eq}(T)\right)
    \label{eq:HSnumber}\\
     \frac{d\rho'}{dt}\!\! &+&\!\! 3  ( 1 + w') H \rho'= - m' \Gamma' \left(  n' -n'_{\rm eq}(T) \right) \, ,
    \label{eq:HSenergytransfer}
\end{eqnarray}
where $x^{(\prime)}=m'/T^{(\prime)}$. {Their derivation, starting from the particle distributions, is detailed in appendix \ref{app:DEs}.\footnote{Similar sets of equations have appeared in the literature, albeit with a different focus, see e.g. \cite{Giudice:1999am,Chu:2011be,Mondino:2020lsc}.} In brief, equation \eqref{eq:continuity} expresses conservation up to expansion of the total energy density, $\rho_{\rm t} = \rho + \rho'$. In that equation, $w_{\rm t} \equiv p_{\rm t}/\rho_{\rm t}$  with $p_{\rm t} = p + p'$ the total pressure is the  equation of the state (eos) of the fluid as a whole. This eos changes smoothly between $w_t = 1/3$ when both sectors are radiation dominated (RD), to $w_t \approx 0$ if the Universe becomes matter dominated by the HS.  Since the visible sector is always RD before BBN, we need only to keep track  of the possible change of the HS eos, $w' = p'/\rho'$, if the dark photons become non-relativistic, see below. Equation \eqref{eq:HSnumber} captures the evolution of the dark photon number density $n'$. The first term on the rhs expresses decay of dark photon from the HS  while the second term represents  production of dark photons from the VS through inverse decay at temperature $T$. The $K_{1,2}$ are modified Bessel functions \cite{Gondolo:1990dk}. They express the temperature dependence of the dark photon decay and inverse decay rates, $\langle \Gamma'\rangle =\Gamma' K_1/K_2$ with $\Gamma'$ the decay rate  in  the dark photon frame, see appendix \ref{app:therm}. At large temperatures, $x^{(\prime)} \ll 1$, $K_1/K_2 \sim {x^{(\prime)}/2}$ due to time dilation. Finally, equation \eqref{eq:HSenergytransfer} captures the evolution of the HS energy density. The first term on the right-hand side corresponds to energy transfer from the HS to the VS, while the second term represents the transfer in the opposite direction, with a VS at temperature $T$. If $\xi \gg 1$, the former is dominant, at least at early times. Both terms involve the HS number density. As explained in appendix \ref{app:DEs}, there is no explicit temperature dependence of the rates, unlike eq.\eqref{eq:HSnumber}. Notice that $m'\Gamma'  n' \sim  (m'/T')\Gamma'  \rho'$ for relativistic dark photons, and so energy transfer is of course also subject to time dilation.  

We will solve these equations as function of the scale factor $a$, using $d/dt = aH(a)d/da$ with $H$ the Hubble rate. As they stand, the equations \eqref{eq:continuity}-\eqref{eq:HSenergytransfer} depend explicitly on $T'$ and $T$. They also depend on  $w'$. In the approximation of MB statistics, $w'$ does not depend on the chemical potential, $w' = p'/\rho'= w'(T')$. We assume that it remains close to its equilibrium value and set $w'\approx w'_{\rm eq} = p'_{\rm eq}/\rho'_{\rm eq}$. This still depends on the pressure in the HS and thus on $T'$. As explained in more details in appendix \ref{app:DEs}, we use a simple but quite accurate approximation for $w'_{\rm eq}$ which depends only on $n'_{\rm eq}(T')$ and $\rho'_{\rm eq}$, cf. equation \eqref{eq:approx_w}. Using this, we finally express the temperature $T'$ as function of $\rho'$ and $n'$, through $T' = p'/n' \approx w'_{\rm eq} \rho'/n'$. Given all these approximations, the  equations \eqref{eq:continuity}-\eqref{eq:HSenergytransfer} form a closed set that can be solved numerically to get $\rho$, $\rho'$ and $n'$ as functions of the scale factor $a$, starting from some initial temperature ratio $\xi_i$ down to the thermalization of both sectors, both for relativistic and non-relativistic dark photons. From such solutions, we can track numerically the evolution of the temperature ratio $\xi = T'/T$ as well as other relevant fluid quantities, like entropy. 
}

\begin{figure}[ht]
    \centering
    \includegraphics[width=0.99\columnwidth]{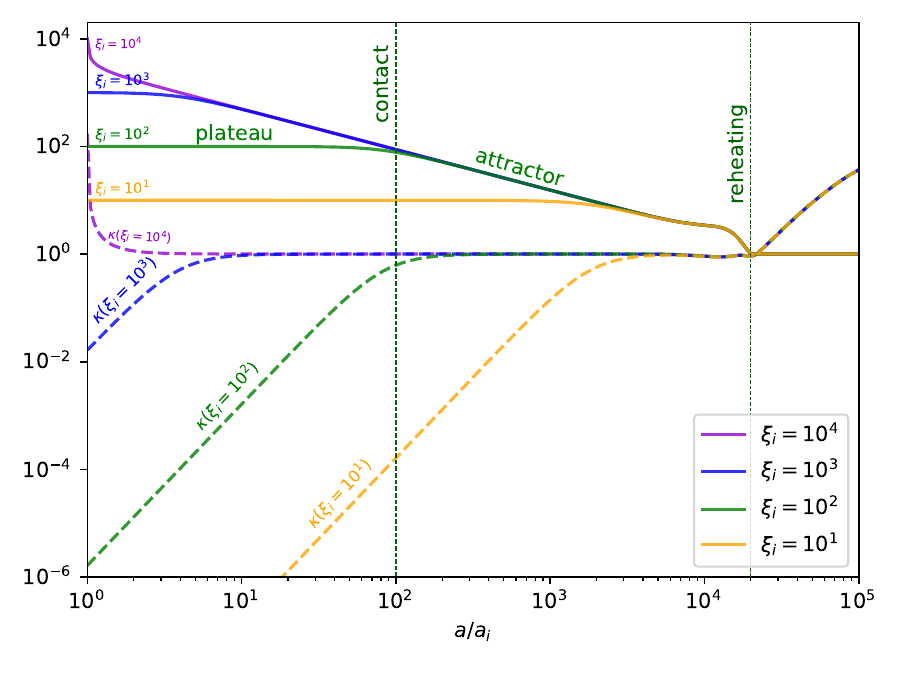}
    \caption{Schematic evolution of the temperature ratio $\xi$ (solid lines) and the heating parameter $\kappa$ (dashed lines) for different $\xi_i$ ($\varepsilon = 10^{-6}$, $m' = 10 ~\rm GeV$). This is illustrated for the case of relativistic reheating, by which we mean that the dark photons are still relativistic when $\xi$ reaches 1. The plateau, attractor and the contact times are indicated for the case of $\xi_i=10^2$. Heating occurs between contact and thermalization. The heating parameter remains constant, $\kappa = {\cal O}(1)$, during heating. See text for details and also fig.\ref{fig:XiSchematic}. {The purple lines (the case of $\xi_i = 10^4$) depict the situation where the temperature ratio is initially unstable (when $\kappa_i \gtrsim 1$).}}
    \label{fig:relativistic}
\end{figure}
\section{Evolution of hidden and visible sectors}
\label{sec:BoltzSolns}

We begin after DM freeze-out, $a_i \sim a_{\rm fo}$. In the following sections, we will discuss in parallel numerical solutions and analytical solutions of the Boltzmann equations \eqref{eq:continuity}-\eqref{eq:HSenergytransfer}.  This will leads us to identify  different  regimes that may occur after DM freeze-out along the process of reheating of the VS as outlined in fig.~\ref{fig:XiSchematic}. We first begin with a discussion of the relevance of the heating parameter.  The, in the following sections, we consider in more details the two possible scenarios alluded to above. First, we discuss relativistic heating, assuming that the dark photons remain relativistic until thermalization of the VS and HS, cf section \ref{sec:rel}. For completeness, we revisit the scenario in which the dark photons become non-relativistic before the end of heating (non-relativistic reheating, see section \ref{sec:nonrel}). Which scenario can be realized in the early universe depends on the  properties of the dark photons, including the kinetic mixing parameter and  is addressed in section \ref{sec:constraints}.

\subsection{Heating parameter}
\label{sec:Heating_param}

\begin{figure*}[t!]
    \centering
 \subfigure{\includegraphics[width=0.48\columnwidth]{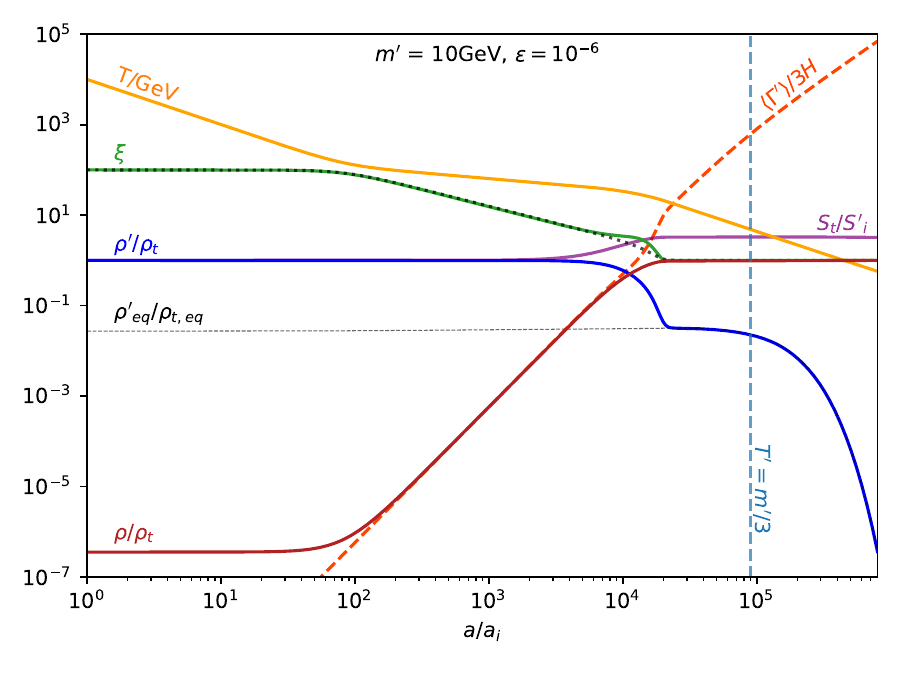}}
    \subfigure{\includegraphics[width=0.48\columnwidth]{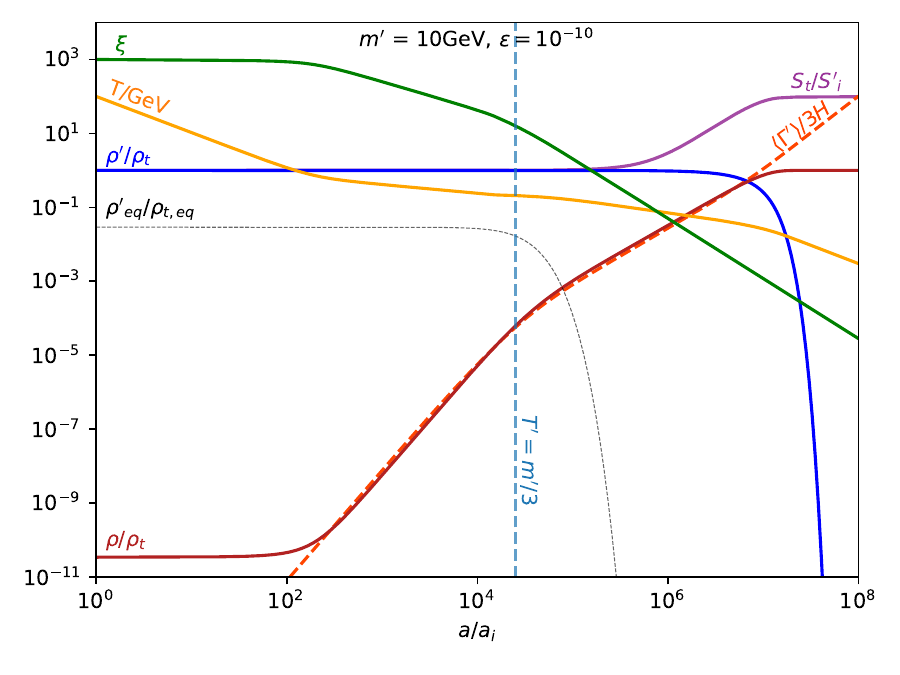}}
    \subfigure{\includegraphics[width=0.48\columnwidth]{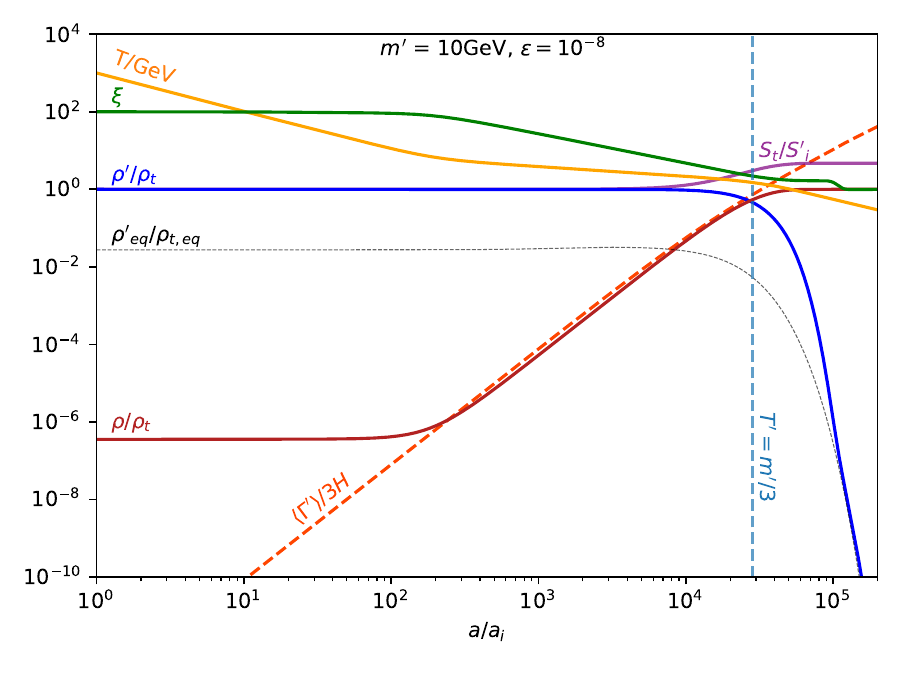}}
    \subfigure{\includegraphics[width=0.48\columnwidth]{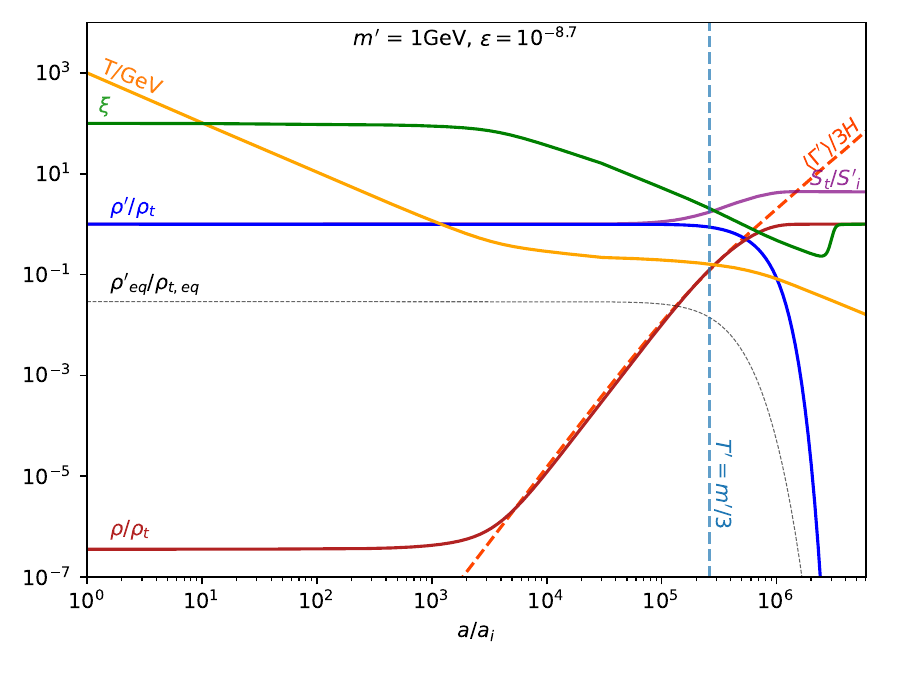}}
    \caption{
    Examples of the different behaviors of the hidden and visible sectors as a function of the scale factor, $a$. 
    The blue, red and thin grey lines give the ratio of energy densities in each sector, $\rho'/\rho_{\rm tot}$ and $\rho/\rho_{\rm tot}$, and the ratio $\rho'_{\rm eq}(T)/\rho_{\rm t, eq}$ respectively. 
    The green, purple, orange and dashed red lines give the temperature ratio, $\xi$, the total entropy production, $S_t/S'_i$, the SM temperature, $T/$GeV, and the thermally-averaged rate, $\langle \Gamma'\rangle/3H$, respectively. 
    The dashed vertical blue lines labeled $T' = m'/3$ indicate approximately when the dark photon becomes non-relativistic. 
    The values of $m'$ and $\varepsilon$ chosen are given in each plot. 
    To avoid unnecessary additional features due to the evolution of $g_\ast$ in the SM, we have fixed $g_\ast = 100$ for these figures. 
    The dotted black line in the top-left panel is the analytic expression for $\xi$, see the comment after eq.\eqref{eq:xiRel}. 
    }
    \label{fig:evolution}
\end{figure*}

{If we assume that the VS and HS  evolve independently at early times, then $T'$ and $T$ both scale as $a^{-1}$. As the temperature ratio $\xi = T'/T$ is constant, we refer to this initial condition as a `plateau', cf fig. \ref{fig:relativistic}. In this section we show that the existence of a plateau is controlled by the following combination of parameters, 
\begin{equation}
\kappa = \frac{1}{3}\frac{\rho'}{\rho}\frac{\langle \Gamma'\rangle}{H}  
\label{eq:heatpar}
\end{equation}
The factor $1/3$ is conventional. We call this combination the 'heating parameter' because efficient energy transfer from the HS to the VS, corresponding to end of the plateau, starts when $\kappa$ becomes order 1 as can be seen in figure \ref{fig:relativistic}. 

This can be understood as follows. We consider a radiation dominated universe  with $T' \gg T$ and both sectors made of relativistic particles. The rhs of eq.\eqref{eq:HSenergytransfer} is dominated by the first term, proportional to $n' \approx n'_{\rm eq}(T')$. Using energy conservation and  $w_t= 1/3$ and setting $\rho' \approx 3 n'T'$, eqs \eqref{eq:continuity} and \eqref{eq:HSenergytransfer} can be combined to give 
\begin{equation}
    {d \ln(a^4 \rho)\over d\ln a}  \approx { \langle\Gamma'\rangle\over 3 H} {\rho'\over \rho}  
    \label{eq:approxEq}
\end{equation}
We see that this equation is sourced by the heating parameter $\kappa$, eq.\eqref{eq:heatpar}. As long as $\kappa \lesssim 1$, $\rho \propto a^{-4}$ and $\xi$ remains constant, corresponding to a plateau. That  $\kappa \lesssim 1$ means that $\langle\Gamma'\rangle/ H \lesssim \rho/\rho'\ll 1$.  Clearly, this condition could not be satisfied if the initial density contrast $\rho'/\rho$ is very large, a  situation we will consider in section \ref{sec:boiling}. Here, instead, we assume that  $\kappa \ll 1$ initially. Then, as $\langle\Gamma'\rangle/ H \propto a^3$,  $\kappa \propto a{^3}$. The moment at which $\kappa \sim 1$ marks the onset of heating of the VS; we call this ``contact", see fig.\ref{fig:relativistic}.  This simply means that a plateau exists as long as the energy of the VS is larger than the one transferred from the HS. This applies whenever there is energy transfer between two sectors, for instance in  freeze-in mechanism, see appendix \ref{app:FI}. 

Perhaps less obvious is that the heating parameter remains ${\cal O}(1)$ throughout the entire heating process, see dashed curves in fig.~\ref{fig:relativistic}. This can be understood as follows. When $\kappa$ reaches $\sim 1$, eq.\eqref{eq:approxEq} implies that $d(a^4\rho)/da \propto a^{2}$ and so $\rho \propto a^{-1}$. Using MB statistics, this implies that $T \propto a^{-1/4}$ and $\xi \propto a^{-3/4}$, see figure \ref{fig:XiSchematic}. Thus, since $\rho' \propto a^{-4}$ and $\langle\Gamma'\rangle/ H \propto a^3$, $\kappa$ remains $\sim 1$. In other words, $\rho \sim \rho'\,\langle\Gamma'\rangle / H$ and the  energy of the VS is entirely determined by the energy transferred from the HS. Starting from plateau-like initial conditions, after thermal contact the temperature ratio $\xi \propto a^{-3/4}$ and becomes independent of the initial temperature ratio, $\xi_i$. The temperature ratio thus follows a solution that behaves as an attractor, see figure \ref{fig:XiSchematic} and also section \ref{sec:attractor}.

The above reasoning assumed that the dark photons are relativistic but remains valid  if they become non-relativistic, as can bee seen in see figures \ref{fig:evolution} which depict some numerical solutions. Indeed, for a Universe dominated by non-relativistic dark photons, $\rho' \propto a^{-3}$ and $\langle\Gamma'\rangle/ H \propto a^{3/2}$ leading to $\rho \propto a^{-3/2}$ during energy transfer. In other words, $\rho \sim \rho'\,\langle\Gamma'\rangle / H$ during heating and this both for relativistic and non-relativistic dark photons. }

\subsection{Relativistic reheating}
\label{sec:rel}

{In this section, we consider in more details the process of relativistic reheating. In particular, we will derive an analytical solution of the Boltzmann equations that closely matches the numerical solutions, describing the evolution of the temperature ratio $\xi = T'/T$ from the onset of heating in the VS to reheating, corresponding to $T' = T$. In particular, we determine analytically at which moments heating starts (``contact") and ends (``reheating").}

\subsubsection{An analytical solution toward reheating}
\label{sec:plateau}

Assuming that dark photons remain relativistic up to the moment of reheating suggests that their abundance can be expressed solely in terms of $T'$, effectively disregarding the chemical potential, $\mu' = 0$. While this assumption is not entirely accurate, as we will explain later, it captures the essence of the relativistic reheating process. Using MB statistics, we can set $w^{\prime} = 1/3$, independent of $\mu'$. Also, $p^{\prime} = n^{\prime} T^{\prime}= g'_\ast T'^4/\pi^2$ and $\rho^{\prime} =\langle E^{\prime}\rangle n^{\prime}$ with  $\langle E^{\prime}\rangle= 3\, T^{\prime}$ setting $\mu' =0$. These relations also hold for the VS, still using MB statistics. 
We track  evolution in terms of the scale factor $a$ instead of time $t$ via $da/dt = aH$. During such RD era, \eqref{eq:continuity} implies that the total energy density scales as $\rho_t \propto a^{-4}$. Using all this, eq.\eqref{eq:HSenergytransfer} can be rewritten as
\begin{equation}
   a H \frac{d}{da}\left(\frac{\rho'}{\rho_t} \right)= -{\Gamma' \over 3 \rho_t} \left[  {m'\over T'}\rho'(T')-{m'\over T} \rho'_{\rm eq}(T) \right] \, .
    \label{eq:HSR}
\end{equation}
{Similarly,  we can write the lhs of eq.~\eqref{eq:HSnumber} in terms of $\rho'$, again using $\rho' = 3 n' T'$ and $dT'/dt = - H T'$. 
On the rhs,  $K_1(x')/K_2(x') \equiv \langle m'/E'\rangle \approx x'/2$ for small $x'= m'/T'$. We thus arrive at the same equation as \eqref{eq:HSR}, but with a slightly more efficient energy transfer rate, as $\langle 1/E'\rangle \gtrsim 1/\langle E'\rangle$. The reason for this difference is straightforward: compared to $\langle E'\rangle$, the moment $\langle 1/E'\rangle$ is biased toward the low-energy part of the dark photon energy distribution. This bias arises because low-energy dark photons decay more quickly than high-energy ones due to time dilation, leading to a departure from equilibrium. This deviation could be captured by including the chemical potential, which we set to $\mu' = 0$. This underscores the need for the complete set of equations \eqref{eq:continuity}-\eqref{eq:HSenergytransfer} to describe in full generality the dynamics of reheating between the two sectors. Moreover, it highlights that a fluid approximation is not entirely suitable for such relativistic particles, as their decay distorts  their distribution. Here, we adopt the way much simpler fluid approximation.
 Addressing the relativistic reheating problem working directly with particle distributions is still work in progress, but preliminary results indicates that the repercussion are faint \cite{HufnagelKT}. To derive an analytical solution, we also disregard the slight difference in the definition of the decay rate and proceed with eq.\eqref{eq:HSR}. As we shall see, the resulting analytical expression is both straightforward and in good agreement with the numerical solutions to the full set of Boltzmann equations.}

We make some last simplifications to derive a useful analytical solution to eq.\eqref{eq:HSR}. First, we notice that the two terms inside the brackets on the rhs of \eqref{eq:HSR} are of order $T^3$ and $T'^3$ respectively. Given that $\xi_i \gg 1$, the dominant term is from decay of the $\gamma'$ $\propto T'^3$, while inverse decay term is not relevant until
$T' \sim T$. 
We therefore set $m'/T \rightarrow m'/T'$ in the last term on the rhs: the numerical difference will be small as long as $\xi \gg 1$, but this change makes the equation easier to handle. 
Then, using $\rho'_{\rm eq}/\rho_{\rm eq} = (g'_\ast /g_\ast) \xi^4$, we can rewrite \eqref{eq:HSR} as an equation for temperatures difference, expressed in terms of
\begin{equation}
    \Delta = T'^4 - T^4 \equiv {\pi^2\over 3}(\rho'/\bar g_\ast  - \rho_t/g_\ast) \, ,
\end{equation}
with $1/\bar g_\ast = 1/g'_\ast + 1/g_\ast \approx 1/g'_\ast$. 
 Thus \eqref{eq:HSR} becomes
\begin{equation}
   a \frac{d}{da}\left(\frac{ \Delta}{\rho_t} \right)\approx - \,\frac{g'_\ast}{\bar g_\ast}\, \frac{m'}{T'} \, {\Gamma' \over 3 H}\,\frac{\Delta}{\rho_t}
    \label{eq:HSRel}
\end{equation}
where we have supposed that $\bar g_\ast$ is constant. Since $g_\ast \gg g'_\ast$ for most of the parameter space we consider, $g'_\ast/\bar g_\ast \approx 1$, so we drop this factor for simplicity.

While the expansion is dominated by the HS,  $T' \propto a^{-1}$, so the solution to \eqref{eq:HSRel} is
\begin{equation}
\label{eq:DeltaEvol}
    {\Delta\over \rho_t}\approx \left.{\Delta\over \rho_t}\right\vert_i e^{-(\sigma -\sigma_i)}
\end{equation}
with $\sigma_i = m'\Gamma'/(9 T'_i H_i)$ and $\sigma = \sigma_i (a/a_i)^3$. 
This leads to a simple expression for the evolution of $\xi$, including thermalization, 
\begin{equation}
    \xi^4 = {(g_\ast'+ g_\ast e^{-(\sigma- \sigma_i)})\xi_i^4 + g_\ast(1 - e^{-(\sigma- \sigma_i)})\over g'_\ast(1 - e^{-(\sigma- \sigma_i)})\xi^4_i + (g_\ast+ g'_\ast e^{-(\sigma- \sigma_i)})}.
    \label{eq:xiRel}
\end{equation}
This expression agrees well with our numerical results. Compare the black dotted  (analytical) and the green solid (numerical) curves in the the upper left panel of fig.~\ref{fig:evolution}. The black dashed line corresponds to the analytical solution, the green solid curve to the the numerical one. There is a small discrepancy between the two curves near $\xi \sim 1$, i.e. thermalization, to which we will return later. From that figure and Eq \eqref{eq:xiRel}, we see that the temperature ratio was constant  at early times, $\xi \approx \xi_i$, then later evolved towards $\xi = 1$. This initial plateau is necessary for the coherence of our scenario, which rests on the hypothesis that the HS and VS each had well defined temperatures at early times.

\subsubsection{Heating from relativistic dark photons}
\label{sec:attractor}
Since $(m'/T') \Gamma'/H \propto T'^{-3}$ increases in time, heat will eventually start transferring from the HS to the VS.
To see the transition of $\xi$ from plateau toward reheating of the VS,  we expand  the exponential to first order. Assuming $g_\ast \gg g'_\ast$ and both $\sigma \ll 1$ and $\xi_i \gg 1$, eq.\eqref{eq:xiRel} is well approximated by
\begin{equation}
    \xi \approx{\xi_i\over \left(1  +\kappa_i((a/a_i)^3 - 1)) \right)^{1/4}}.
      \label{eq:xiRelapp}
\end{equation}
This expression involves the initial value of a combination of parameters that we identify as  the heating parameter $\kappa$ defined in eq.\eqref{eq:heatpar},
\begin{equation}
\kappa_i  = \frac{\rho'_i}{3 \rho_i}\frac{\langle \Gamma'\rangle_i}{H_i}.
\label{eq:kappa_def}
\end{equation}
While we assumed $\sigma_i \lesssim 1$ to derive \eqref{eq:xiRelapp}, $\kappa_i$ can be smaller or larger than 1, depending on the initial temperature ratio $\xi_i$. From inspection of \eqref{eq:xiRelapp}, we expect that a plateau exists at early times provided $\kappa_i \lesssim 1$. The precise condition turns out to be $\kappa_i < 4/3$, see eq.\eqref{eq:Tmax1} and section \ref{sec:boiling} where, for completeness, we discuss the case $\kappa_i \gtrsim 1$. 

Assuming $\kappa_i \ll 1$, the temperature ratio stays constant while $\kappa = \kappa_i (a/a_i)^3$. It eventually reaches $\kappa_c \approx 1$, after which $\xi$ starts to decrease, marking the onset of energy transfer from the HS to the VS. We call this event `contact', see fig.\ref{fig:relativistic}. 
From \eqref{eq:xiRelapp}, contact occurred at
\begin{equation}
   \frac{a_{c,i}}{a_i} \simeq {1\over \kappa_i^{1/3}}  \sim \left(\frac{g_*}{\sqrt{g'_*}} \frac{ T_i'^3}{m'\Gamma' m_{\rm Pl}} \right)^{1/3} {1\over \xi_i^{4/3}} \,.
   \label{eq:contact}
\end{equation}
Here and below, the subscript $i$ is meant to emphasize the fact that $a_{c,i}$ depends on $\xi_i$, with contact occurring later for smaller values of $\xi_i$. 
From eq.(\ref{eq:contact}), and using that both $T$ and $T'$ evolve as $1/a$ before contact, the temperature of the VS at contact is
\begin{equation}
    T_{c,i} \approx {0.4} \left(\frac{g'_\ast}{g_\ast}\right)^{1/3} \left( \frac{m' \Gamma' m_{\rm Pl}}{g_\ast{'}^{1/2}} \right)^{1/3} \xi_i^{1/3}\,.
    \label{eq:contact_temperature}
\end{equation}
Also, for $a \gtrsim a_c$, the temperature ratio evolves as
\begin{align}
    \xi  \approx \xi_i  \left(  \frac{ a_{c,i}}{a} \right)^{3/4} \approx  \left( \frac{ g_\ast}{g'_\ast \sigma_i }  \right)^{1/4} \left(  \frac{ a_i}{a} \right)^{3/4}.
    \label{eq:attractor}
\end{align}
From \eqref{eq:contact} and \eqref{eq:attractor}, we see that  $\xi$ is independent of $\xi_i$ after $a_{c,i}$. So this part of its evolution is an attractor, see solid curves in fig.\ref{fig:relativistic}.

While $\kappa$ increases before contact, it remains constant during heating, $\kappa \approx 1$, see dashed curves in fig.\ref{fig:relativistic}. Indeed, from \eqref{eq:attractor}, $\rho'/\rho \sim \xi^4 \propto a^{-3}$ while $\langle \Gamma'\rangle/H \propto a^3$. Thus, the energy of the VS evolves as $\rho \approx \rho' \langle \Gamma'\rangle/H$, see section \ref{sec:Heating_param}. For $\xi_i \gg 1$, contact and subsequent heating occur while the decay rate is still (much) less than the Hubble rate, $\langle \Gamma'\rangle/H \ll 1$. Clearly, for  $\xi_i \gg 1$, there is a large reservoir of energy that can be transferred from the HS to the VS, hence the relevance of the ratio of rates and energy densities in the heating parameter. 

Since $T' \propto a^{-1}$ to a good approximation when $\xi_i \gg 1$, the temperature of the VS evolves as
\begin{equation}
 {T\over T_i} \approx \left({a_{c,i}\over a}\right)^{1/4} \propto t^{-1/8} \quad \mbox{\rm and} \quad\xi \propto a^{-3/4},
 \label{eq:Tattractor}
\end{equation}
see e.g. the orange solid curve in the upper left panel of fig.\ref{fig:evolution}.
The temperature of the VS  decreases but quite slowly during heating, a behaviour that is qualitatively similar but quantitatively distinct from the case of heating through the decay of a non-relativistic particle, see eq.\eqref{eq:TST} and the summary of the comparison between relativistic and non-relativistic thermalization in fig.\ref{fig:XiSchematic}.

\subsubsection{Thermalization of dark photons and aftermath}
\label{sec:rel_eq}
According to eq.\eqref{eq:xiRel}, $\xi$ decreases until the VS and HS reach thermalization. Naively, the condition for this to occur is $\langle \Gamma'\rangle/H \gtrsim 1$.
However, numerical solutions (see fig.\ref{fig:evolution}) and approximate analytical solution \eqref{eq:attractor} reveal that the reheating condition depends on the relative number of degrees of freedom between the HS and VS. Indeed, $\kappa \approx 1$ at $\xi \approx 1$, corresponding to 
$ \langle \Gamma'\rangle/H \sim g'_\ast/g_\ast$, see dashed red curves in fig.\ref{fig:evolution}. 
Once the sectors thermalize, $\kappa$ increases again, $\kappa \propto \langle \Gamma'\rangle/H$. The heating parameter thus provides a simple criteria to track the energy transfer between the sectors. Put simply, $\kappa < \mathcal{O}(1)$ before heating, $\approx 1$ during heating, and $> \mathcal{O}(1)$ after reheating (see Figs. \ref{fig:XiSchematic} and \ref{fig:relativistic}).

Using eq.\eqref{eq:attractor} to estimate the moment of reheating, gives $a_{\rm th} \approx a_{c,} \xi_i^{4/3}$ and 
\begin{equation}
    T_{\rm rh} \approx {0.4} \left(\frac{g'_\ast}{g_\ast}\right)^{1/3} \left( \frac{m' \Gamma' m_{\rm Pl}}{g_\ast{'}^{1/2}} \right)^{1/3}.
    \label{eq:T_rh}
\end{equation}
\begin{figure}[t]
    \centering
    \includegraphics[width=0.99\columnwidth]{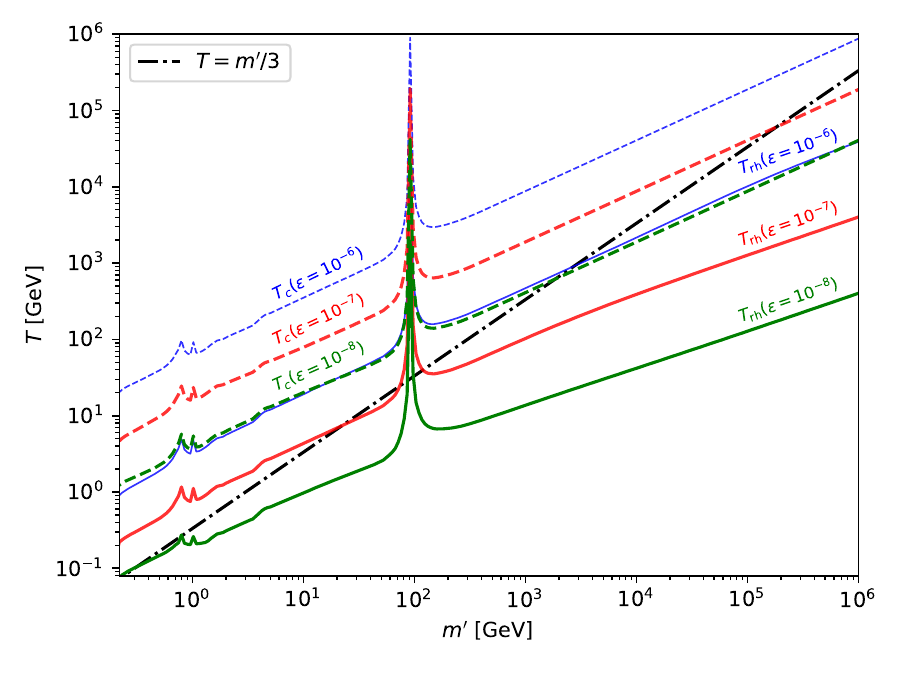}
    \caption{Reheating (solid) and contact (dashed) temperatures in function of $m'$ for different values of $\varepsilon$. Contact temperature is given for $\xi_i = 10^3$. Above (below) the dot-dashed line, the dark photons are relativistic (resp. non-relativistic) at the time of reheating. The features are due to mixing between the $\gamma'$ and resonances in the SM photon channel, $J/\psi,\ldots, Z$. The $\varepsilon=10^{-6}$ case is the maximal admissible value of the mixing parameter if we take into account the requirement of DM freeze-out {in a secluded HS} with $\xi_i \gtrsim 1$, as explained in section \ref{sec:constraints}.}
    \label{fig:Tth}
\end{figure}
As expected from the attractor behavior of the evolution of $\xi$,  $T_{\rm rh}$  is independent of $\xi_i$. 
Incidentally, \eqref{eq:T_rh} is precisely the same as \eqref{eq:contact_temperature} if we set  $\xi = 1$.\footnote{The numerical pre-factor, close to the value $0.34$ obtained using \eqref{eq:attractor}, has been chosen to closely match the result  from the numerical determinations  of $T_{\rm rh}$, as depicted in fig.~\ref{fig:Tth}.} As expected, $T_{\rm rh}$ is equivalent to the reheating temperature  in the more familiar case of decay of a non-relativistic particle {with a decay rate $\Gamma$}, 
$T_{\rm rh} \sim \sqrt{\Gamma m_{\rm Pl}/g_\ast^{1/2}}$, see \cite{Kolb:1990vq} and section \ref{sec:nonrel}. Indeed, replacing $\Gamma$ in the latter  by the decay rate of a relativistic particle, $\sim (m/T)  \Gamma$ gives back
(\ref{eq:T_rh}). The difference is that the HS and VS are made of relativistic degrees of freedom all along.

The entire discussion above assumes that the dark photons are relativistic at $T_{\rm rh}$. For a given $\gamma'$ decay rate, the condition that $T_{\rm rh} \gtrsim m'/3$, together with the expression for the reheating temperature \eqref{eq:T_rh}, imposes that 
\begin{equation}
m' \lesssim \left(\frac{g'_\ast}{g_\ast}\right)^{1/2} \left({m_{\rm Pl} \Gamma'\over g_\ast^{\prime 1/2}}\right)^{1/2}.
\label{eq:Tth}
\end{equation}
As $\Gamma' \propto \epsilon^2 m'$, this sets an upper bound on $m'$ for a fixed mixing parameter. 
Constraints on the mixing parameter and $\gamma'$ mass will be discussed in section \ref{sec:constraints}. In the meantime, fig.\ref{fig:Tth} illustrates how the reheating (solid) and contact (dashed) temperatures depend on the $\gamma'$ mass and mixing parameter for $\xi_i = 10^3$ (see also caption). One possible use of this figure may be the following. Consider for instance $\varepsilon = 10^{-7}$ and $m' = 20$ GeV. We see that contact occurred when the temperature of the VS was $T_c \approx 100$ GeV and reheating at $T_{\rm rh} \approx 6$ GeV, when the $\gamma'$ was becoming non-relativistic (see dot-dashed line). This implies, that the electroweak phase transition could have occurred along the plateau, with an expansion of the Universe driven by a (much) hotter HS \cite{Fairbairn:2019xog}.

We note that after relativistic reheating, the dark photons remain in thermal equilibrium, even when they become non-relativistic. Unlike annihilation processes, decay and inverse decay processes both remain effective at late time. This is clear for decay, as $\Gamma' > H$ always after reheating. Inverse decay is proportional to the $\gamma'$ equilibrium abundance, which implies that the $\gamma'$ always tracks their equilibrium abundance \cite{Harvey:1981yk,Kolb:1990vq}, see appendix \ref{app:FI}. This can be seen in the solid blue curves in fig.\ref{fig:evolution}, which depicts the evolution of $\rho'/\rho_t$. In the upper left panel, corresponding to relativistic reheating, $\rho'/\rho_t$ first drops to $\rho'/\rho_t \approx g'_\ast/g_\ast$ and then becomes Boltzmann suppressed with $\rho' \approx m' n'_{\rm eq}(T)$ when the dark photons are non-relativistic; the equilibrium abundance at $T$ is depicted as the light dotted curve. The dark photons must thus be heavier than $m' \gtrsim 5$ MeV. Otherwise, their relative abundance is too large to fulfill the $\Delta N_{\rm eff}$ condition (see section \ref{sec:constraints}).

The other panels correspond to non-relativistic reheating. In particular, in the lower left panel the dark photons are initially over-abundant when they become NR but eventually track their equilibrium, $\rho' \approx m' n'_{\rm eq}(T)$. For the choice of parameters depicted in the two right panels, inverse-decay processes are essentially irrelevant and the evolution of the $\gamma'$ is entirely dominated by direct decay down to very low, and thus negligible, $\gamma'$ abundances.

\subsubsection{Relativistic entropy production}
\label{sec:entropy_rel}

The heat transfer from the hidden to the visible sector is an irreversible process, hence entropy production is expected. 
A direct calculation of entropy evolution is given in appendix \ref{app:entropyProduction}. The outcome is as follows. For relativistic reheating, 
\begin{equation}
    S_{t,f}/S_{t,i} \approx (g_\ast/g_\ast')^{1/4}
    \label{eq:entropyHHSRD}
\end{equation}
where $S_{t,f}$ ($S_{t,i}$) is the final (initial) total comoving entropy. This expression assumes $\rho'_i \gg \rho_i$, so that the initial entropy was dominantly within the HS, $S_{t,i}\approx S_i'$, while $S_{t,f}\approx S_f$ if $g_\ast \gg g_\ast'$, see eq.\eqref{eq:entropyProdRel} \cite{Coy:2021ann}. In appendix \ref{app:entropyProduction}, we show that the VS entropy increases as $S \propto a^{9/4}$, slightly faster than in the case  of non-relativistic decay, $S \propto a^{15/8}$ \cite{Scherrer:1984fd}.  

For our numerical calculations, entropy is calculated directly from the evolution of the temperatures $T$ and $T'$ using the MB equilibrium expression for the entropy (see appendix \ref{app:therm}). We checked that our numerical solutions match the asymptotic analytical expressions. An instance of the evolution of the entropy produced in the case of relativistic reheating is depicted by the purple line in the upper left panel of fig.\ref{fig:evolution}.

For $g_\ast \approx 100$ and a massive $\gamma'$, the entropy per comoving volume has increased by a factor $\sim 3$. After reheating, the DM abundance changed from \eqref{eq:DMabundance} to
\begin{equation}
Y_{\rm dm} =  \frac{S_{t,i}}{S_{t,f}} Y_{\rm dm}' \propto\left(\frac{g_\ast'}{g_\ast}\right)^{1/4} \!{x'_\text{fo}\over g'^{1/2} m_{\rm dm} m_{\rm Pl} \langle \sigma v \rangle}.
\label{eq:DMdilution}
\end{equation}
Entropy dilution in the scenario of relativistic thermalization is modest but non-negligible, being ${\cal O}(1)$. It is also irreducible,  as an even larger entropy can be produced  if the $\gamma'$ become non-relativistic before thermalization, see section \ref{sec:nonrel}.

\subsubsection{Relativistic maximal temperature $T_{\rm max}$}
\label{sec:boiling}
Before closing the section on relativistic reheating, we comment on the possibility of a large initial heating parameter, $\kappa_i \gtrsim 1$. 
We have seen that $\kappa_i \lesssim 1$ implies that $\xi \approx \xi_i$, that both the visible and hidden sector temperatures evolve as $a^{-1}$, and that energy transfer from HS to VS is negligible compared to the amount of energy already present in the VS. A similar condition holds, with exchange $\rho \leftrightarrow \rho'$, if instead the VS is the dominant sector, as in standard freeze-in scenarios. In that case we must make sure that the Freeze-in contribution does not supersede the initial condition, so that our $\xi_i$ parameter defines well the situation before the start of thermalization. This question is treated in appendix \ref{app:FI}.

Now consider $\kappa_i \gtrsim 1$. 
This could correspond to two distinct situations, depending on whether $\langle \Gamma'\rangle/H$ is smaller or larger than 1. The latter would imply that reheating is instantaneous or, i.e. that the HS and VS should immediately be in equilibrium. This contradicts our working hypothesis, so we will use it to put a bound in the kinetic mixing parameter vs. $m'$ mass plane in section \ref{sec:constraints}. The situation in which initially $\langle \Gamma'\rangle/H \lesssim 1$ while $\kappa_i \gtrsim 1$ is more interesting. This corresponds to a very large energy density ratio. Instead of a plateau, the VS temperature first rises to a maximal temperature, $T_{\rm max} \gtrsim T_{\rm rh}$. 
To see this, we employ eq.\eqref{eq:xiRelapp} which, using that $T'\propto a^{-1}$, can be rewritten as 
\begin{align}
    {T\over T_i} = {a_i\over a}\left(1  + \kappa_i \left((a/a_i)^3 - 1\right) \right)^{1/4}.
    \label{eq:Tmax1}
\end{align}
From this, one can check that instead of decreasing, $T$ first rises to reach a maximum $T_{\rm max}$ 
at 
\begin{equation}
\left({a_{\rm max}\over a_i}\right)^3 = 4 \left(1 - 1/\kappa_i\right)> 1.
\end{equation}
This is provided $\kappa_i > 4/3$, which is the more precise condition for the absence of a plateau.
If $\kappa_i \gg 1$, $a_{\rm max} \approx 2^{2/3} a_i$, so heating to $T_{\rm max}$ is very rapid. As $T' \propto a^{-1}$, for $\kappa_i \gg 1$ this corresponds to 
\begin{equation}
\label{eq:tmax}
   T_{\rm max} \approx \frac{(3 \kappa_i)^{1/4}}{4^{1/3}} T_i \approx 0.6 g_\ast^{-1/4} \left( \langle\Gamma'\rangle_i m_{\rm Pl} \right)^{1/4}\,(g'_\ast T^{\prime 4}_i)^{1/8}\,.   
\end{equation}
{Notice that this temperature of the VS is maximal only in the sense that $T$ will decrease after having reached $T_{\rm max}$. It also acts a lower bound for the temperature $T$ for a given set of parameters $m_{\rm dm}$, $m'$ and $\varepsilon$. Indeed, cases for which $T_i > T_{\rm max}$ correspond to smaller initial values of the heating parameter $\kappa$, for which the temperature ratio has a plateau at early times. Thus, in general $T_{\rm max} < T < T'$ when the hidden sector is initially dominant.}
The temperature ratio at $a_{\rm max}$ is
\begin{equation}
    \xi(a_{\rm max}) \approx {\xi_i \over \left( 3 \kappa_i  \right)^{1/4}} < \xi_i 
\end{equation}
which is independent of $\xi_i$ since $\kappa_i \propto \xi_i^4$. As $T$ reaches $T_{\rm max}$, the heating parameter rapidly decreases toward $\kappa \sim 1$ and stays so until reheating (see the dashed purple curve in fig.\ref{fig:relativistic}).

A rapid increase of the VS temperature is at odds with our basic assumption, namely that the hidden and visible sectors are effectively decoupled around DM freeze-out. Later on, we will use the condition $\kappa_i \lesssim 1$ to set constraints on the possible initial temperature ratio $\xi_i$ in the domain of thermal DM candidates, see sections \ref{sec:constraints} and \ref{sec:domain}. 

This is all very similar to the standard inflationary scenario. In that case, the HS consists only of the inflaton and the VS is essentially at zero temperature so that reheating to $T_{\rm rh}$ is preceded by a rapid rise to a maximal temperature $T_{\rm max} > T_{\rm rh}$ \cite{Kolb:1990vq,Giudice:1999am}. Eq.\eqref{eq:tmax} can be directly compared with the corresponding expression in the case of inflaton decay with mass $M_I$, initial abundance $\rho_I$ and decay rate $\Gamma_I$,
\begin{equation}
   T_{\rm max} \sim g_\ast^{-1/4} \left(\Gamma_I m_{\rm Pl} \right)^{1/4} \rho_I^{1/8}
   \label{eq:T_max_inflation}
\end{equation}
see eq.(8.33) in \cite{Kolb:1990vq} with $\rho_I \approx M_I^4$  (see also \cite{Giudice:1999am}). 
The equivalence between the two cases amounts to replacing $ \Gamma_I$ by $\langle\Gamma'\rangle_i$ and $\rho_I$ by $\rho_i' \sim g'_\ast T_i'^4$. 
\footnote{{In expression \eqref{eq:T_max_inflation}, $T_{\rm max} \rightarrow T'_{\rm max}$ if the inflaton decays dominantly into the HS. This raises the question of whether it may be possible to set an upper bound on the temperature of the HS and VS sectors? In the scenario of a hot HS, it is evident that $T \lesssim T'$. The relevant upper bound is thus on \( T' \). With the bound on the inflationary scale \( H_I \lesssim 10^{-5} m_{\rm pl} \) \cite{Planck:2018nkj}, assuming \( \Gamma_I \lesssim H_I \sim \rho_I^{1/2}/m_{\rm pl} \),  \eqref{eq:T_max_inflation} gives that 
\( T'_i\lesssim 10^{17} \, \text{GeV}\) in \eqref{eq:T_max_inflation} . The interplay between reheating from inflation and the subsequent reheating of the VS from a hot HS is work in progress \cite{Clery:XXXX}.}
}

\subsection{Non-relativistic reheating}
\label{sec:nonrel}
In the previous section, we considered the possibility that reheating of the VS occurred when the dark photons are still relativistic, $T_{\rm rh} \gtrsim m'/3$. As expected and as fig.\ref{fig:Tth} shows, heavier and/or longer-lived dark photons tend to become non-relativistic when the HS is still hotter than the VS, at which point the Universe becomes matter-dominated. The key characteristics of this non-relativistic reheating scenario are given in the lower part of fig.~\ref{fig:XiSchematic}. Typically, the evolution of the temperature ratio starts with a plateau, followed by an early phase of relativistic heating, until the dark photons become NR. The problem of heating through the decay of a non-relativistic particle is textbook  \cite{Scherrer:1984fd,Kolb:1990vq}. However, some specific features in our scenario are less standard. In particular, in this section we study the transition from the relativistic to the non-relativistic regime and how this information can be used to determine the entropy dilution factor.

\subsubsection{Heating from non-relativistic dark photons}
\label{sec:ST}
We consider dark photons that becomes non-relativistic while $T' \gg T$. Using MB statistics, this occurs roughly when  $T' = T'_{\rm nr} \approx m'/3$. Assuming first a phase of relativistic heating, this corresponds to $a_{\rm nr} \simeq 3 T'_i a_i/m'$ with  $\xi_{\rm nr}$ given by \eqref{eq:attractor}. At that moment, the dark photon decay rate is still smaller than the expansion rate and, as we will see below, the temperature of the VS evolves as $a^{-3/8}$, see fig.\ref{fig:XiSchematic}. We  refer to this regime as non-relativistic heating (or reheating depending on the context) \cite{Scherrer:1984fd}.

For non-relativistic dark photons, Eqs.\eqref{eq:HSnumber} and \eqref{eq:HSenergytransfer} become essentially equivalent, since $\rho' \approx m' n'$. 
Neglecting dark photon inverse decays, since the HS dominates, the equations reduce to
\begin{equation}
    \frac{dn'}{dt} + 3 H n'\approx - \Gamma' n'\, ,
\end{equation}
with solution
\begin{equation}
    n' a^3= a_{\rm nr}^3 n_{\rm nr}'\exp \left(- \frac{2\Gamma'}{3 H_{\rm nr}}((a/a_{\rm nr})^{3/2}-1) \right)
    \label{eq:npNR}
\end{equation}
using that the expansion is MD, and following the evolution from when the dark photons became non-relativistic. 
From Eq.~\eqref{eq:continuity}, 
the VS energy density evolves as
\begin{equation}
    \frac{d}{da}(\rho a^4) = \frac{m' \Gamma'n'a^3}{H }.
    \label{eq:rhoBE}
\end{equation}
The non-relativistic heating period corresponds to the early heating of the VS, when $\Gamma' \lesssim H$, in which case the exponential in \eqref{eq:npNR} is close to 1. In this approximation, integrating \eqref{eq:rhoBE}  gives
\begin{equation}
    \rho = \rho_{\rm nr}\left(\frac{a_{\rm nr}}{a}\right)^4 + \frac{2}{5} \frac{\Gamma'}{H_{\rm nr}} \rho_{\rm nr}^\prime\left(\frac{a_{\rm nr}}{a}\right)^4 \left(\left(\frac{a}{a_{\rm nr}}\right)^{5\over2} - 1 \right)
    \label{eq:rhoST}
\end{equation}
which we can rewrite as an equation for the evolution of $T$ for $\Gamma' \lesssim H$,
\begin{equation}
    {T\over T_{\rm nr}} = \frac{a_{\rm nr}}{a}\left(1+ {6\over 5} \kappa_{\rm nr} \left((a/a_{\rm nr})^{5/2} - 1\right)\right)^{1/4} 
    \label{eq:heatingNR}
\end{equation}
where
\begin{equation}
\kappa_{\rm nr} = {\rho'_{\rm nr}\over 3 \rho_{\rm nr}} \, { \Gamma'\over  H_{\rm nr}}.
\label{eq:kappanr}
\end{equation}
Eq.\eqref{eq:heatingNR} is the counterpart of \eqref{eq:Tmax1}. The factor $6/5 \approx 1$ is there because the expansion is MD instead of RD and $\langle \Gamma'\rangle = \Gamma'$. Otherwise, the features are the same as for \eqref{eq:Tmax1}. For small $\kappa_i \ll 1$, $\kappa \propto a^{5/2}$, instead of $\propto a^3$ in the case of relativistic dark photon, while $\kappa = {\cal O}(1)$ during the early phase of heating, as $\rho \approx \rho' \Gamma'/H$.

Here, we want to track the evolution of  $T$ (and $\xi$) starting  with relativistic dark photons, see fig.~\ref{fig:XiSchematic}. The heating parameter $\kappa_{\rm nr} = {\cal O}(1)$ when the $\gamma'$ particles become non-relativistic, and
\begin{equation}
 {T\over T_{\rm nr}} \approx \left({a_{\rm nr}\over a}\right)^{3/8} \propto t^{-1/4} 
 \label{eq:TST}
\end{equation} 
for $a \gtrsim a_{\rm nr}$.
Since the dark photons are non-relativistic and indeed decoupled, their temperature (or more precisely, mean energy) scales as $T' \propto a^{-2}$, so that $\xi \propto a^{-13/8}$. An instance of such evolution is depicted in the upper right panel of fig.\ref{fig:evolution}, see the green and orange curves for $\xi$ and $T$. See also $\rho/\rho_t \approx \rho/\rho'$ (solid red) and $\langle \Gamma'\rangle/3 H$ (dashed red) which imply that $\kappa = \rho'/\rho \langle \Gamma'\rangle/3 H \sim 1$ during both the phases of relativistic and non-relativistic heating.

\subsubsection{Non-relativistic entropy production}
\label{sec:entropy_nonrel}

From eq.\eqref{eq:rhoST}, we see that the VS energy density evolves as $\rho \sim \rho' \Gamma'/H$ until $\Gamma' \sim H$, at which point the $\gamma'$ particles have shed most of their energy into the VS. The entropy produced can be derived from the evolution of the  temperature $T$ using the MB relation $s = 4 \rho/T$. This implies that the VS comoving entropy evolves as $S \propto a^{15/8}$. 
Alternatively, we can directly solve a Boltzmann equation for entropy evolution \cite{Kolb:1990vq,Scherrer:1984fd}. As the situation we consider is non-standard, we revisit this in appendix \ref{app:entropyProduction}. In the present section, we just quote key results. 

Considering  dark photons that became non-relativistic at $a_{\rm nr}$, the final entropy after their decay is given by eq.\eqref{eq:entropyHHS}, 
\begin{equation}
    {S_{t,f}\over S_{t,i}} \approx {S_f\over S_i'} \approx \left({g_\ast\over g_\ast'}\right)^{1/4}  \left({\tau'\over t_{\rm nr}}\right)^{1/2}
    \label{eq:entropyNR1}
\end{equation}
with $\tau' = 1/\Gamma' \gtrsim t_{\rm nr}$. 
The inverse of this factor leads to entropy dilution of the initial DM abundance, as in eq.\eqref{eq:DMdilution}. Eq.\eqref{eq:entropyNR1} stems from the more general expression \eqref{eq:entropyST1}, assuming $\tau' \gtrsim  t_{\rm nr}$, using MB statistics to relate entropy to energy densities, and using the fact that before the dark photons are non-relativistic, the entropy is  dominantly within the HS.
We see that the more stable the $\gamma'$, the larger is the entropy produced, with a ratio that grows as the square root of the $\gamma'$ lifetime. If $\tau'\sim t_{\rm nr}$, we recover the entropy produced in the case of relativistic heating, eq.\eqref{eq:entropyProdRel}. We have checked that \eqref{eq:entropyNR1} is in very good agreement with our numerical calculations. 

Note that thermalization does not proceed as simply as in the relativistic case (section \ref{sec:boiling}). In the NR case, the entropy produced reaches a maximum  roughly when $\rho' \approx \rho$ and the expansion is again RD, dominated by the VS. The temperature of the VS at that moment is the standard heating temperature 
\begin{equation}
    T_{\rm rh} \sim \left({m_{\rm Pl} \Gamma'\over g_\ast^{1/2}}\right)^{1/2}
    \label{eq:teqNR}
\end{equation}
after which  $T \propto a^{-1}$ \cite{Scherrer:1984fd,Kolb:1990vq}. 
The $\gamma'$ particles are subdominant and  will keep on decaying. If at some moment $n' \propto e^{-t/\tau'} \sim e^{-m'/T}$, then, due to inverse decay processes,  the dark photons will follow an equilibrium abundance at the temperature of the VS, $n' = n_{\rm eq}(T)$, see \cite{Harvey:1981yk} and appendix \ref{app:FI}. We refer to this as the equivalent of thermalization in the non-relativistic scenario, see fig.\ref{fig:XiSchematic}. An instance of such evolution is depicted in the lower right panel of fig.\ref{fig:evolution}. Note that $T'$, which was decreasing as $a^{-2}$, characteristic of NR particles, rises toward $T$ after thermalization occurs. 
In the top right panel, we depict a case for which thermalization never takes place in practice, in the sense that inverse decay processes are negligible down to extremely low $\gamma'$ number densities. This is evident from the difference between the blue `true' and dashed grey `equilibrium' ratios for $\rho'/\rho_{\rm t}$.

\section{Implications for dark photon mass and mixing}
\label{sec:constraints}
\begin{figure}[t!]
    \centering
    \includegraphics[width=0.99\columnwidth]{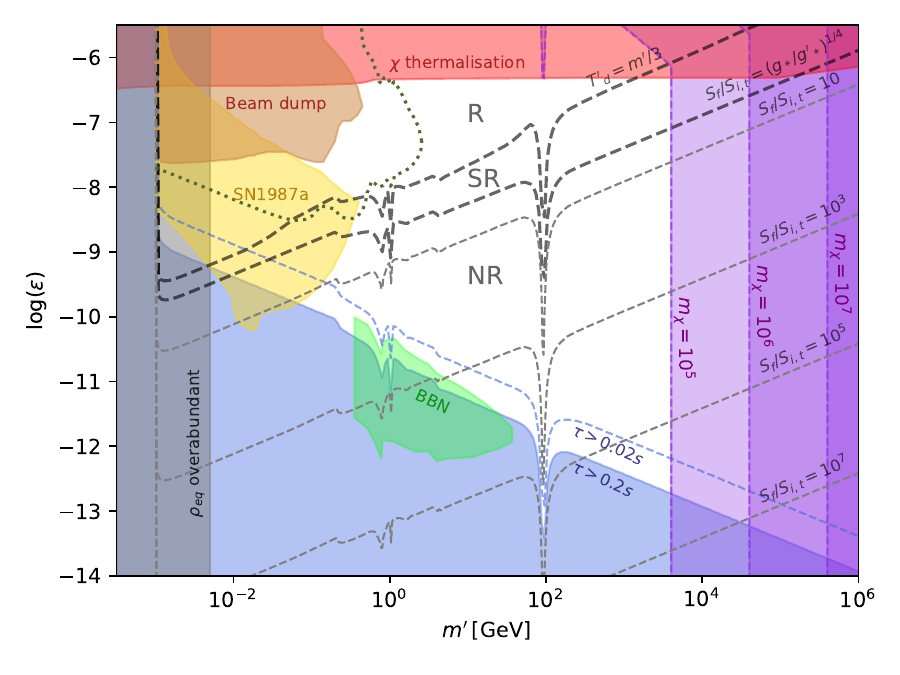}
    \caption{Dark photon parameter space in the scenario with an initially dominant dark sector ($\xi_i \gg 1$). Constraints include present (brown) and expected future (green dotted) beam dump experiments \cite{Bauer:2018onh}, supernovae (yellow) \cite{Hardy:2016kme} and the BBN bound computed in \cite{Berger:2016vxi} (bright green). 
    Candidates in the blue and grey regions are  overabundant at $T = \rm 1$ MeV ($\rho'/\rho_t > 0.04$). DM has thermalized with the VS for the candidates in the red region. In the purple region (whose boundary depends on the DM mass, $m_{\rm dm}$), the dark photons have thermalized before the DM freeze-out.}
    \label{fig:constraints}
\end{figure}

We now use the results from the previous sections to delineate a domain in the  $\varepsilon - m'$ parameter space consistent with our scenario, see fig.~\ref{fig:constraints}. The implications for the domain of the DM particle $\chi$ are discussed in the next section. 

\subsection{Imposing DM freeze-out in the hidden sector}
\label{sec:nothermbeforefo}

The cosmological scenario we consider essentially rests on two hypotheses. First, we assumed that the Universe was dominated by a hot HS, $\xi \gg 1$, before BBN. Second, we assumed that, during that period, the DM abundance was set by secluded thermal freeze-out, through $\chi \bar \chi \rightarrow \gamma'\gamma'$ processes. Concretely, we require that the hidden and visible sectors were effectively decoupled so that $\xi \approx$ constant around DM freeze-out. The processes that could lead to the thermalization of the two sectors are $\gamma'$ decay and inverse decay on one hand, and DM annihilation (production) into (from) VS particles on the other hand. We begin with the latter. {The relevant cross section and decay rate are given in appendix \ref{app:cross_sec}.}

Considering annihilation, we consider $\chi \bar \chi \leftrightarrow f \bar f$ and require that these processes were out of equilibrium at $T' \approx m_{\rm dm}$, here the mass of the $\chi$ particle. That moment gives the strongest constraint because the annihilation rate $\Gamma_{f\bar f}/H {\equiv n_{\rm dm} \left< \sigma_{\chi \bar \chi \rightarrow f \bar f} v \right>/H} \propto 1/T' \sim a$ when the DM particles are relativistic, but is Boltzmann suppressed when they become non-relativistic, so the maximum lies around $T' \approx m_{\rm dm}$ \cite{Coy:2021ann,Arcadi:2019oxh}. 
The condition that the two sectors do not equilibrate at that moment is  ${\Gamma_{f\bar f}(m_{\rm dm})} \lesssim H(m_{\rm dm})$ or
\begin{equation}
\varepsilon \lesssim  \left({m_{\rm dm}\over \alpha' m_{\rm Pl}}\right)^{1/2} \quad \mbox{\rm (no $\chi$ thermalization)}
\label{eq:notherm}
\end{equation}
As we assumed that DM abundance is set by secluded freeze-out of $\chi$ into dark photons, $\alpha'^2/m_{\rm dm}^2\sim  \langle \sigma v\rangle \sim $ pbarn can be traded for the DM abundance (see below for the possible effects of entropy dilution).
This leads to an upper bound on the mixing parameter of $\varepsilon \lesssim  10^{-6}$, corresponding to the red region in fig.~\ref{fig:constraints}. The simple estimate neglects the fact that $\gamma'$ decays produce entropy and thus dilutes the DM abundance, but this effect (including in the relativistic regime) is taken into account in fig.~\ref{fig:constraints}. As smaller values of $\alpha'/m_{\rm dm}$ are required (weaker annihilation cross section), the bound on $\varepsilon$ is weaker. For $\varepsilon \sim 10^{-6}$, the effect becomes  relevant for $m' \gtrsim 10^5$ GeV and leads to a slight rise in the bound on $\varepsilon$. 

Next we require that the dark photons have not thermalized with the VS at the time of DM freeze-out, $x'_{\rm fo} = m'/T'_{\rm fo}$, so that freeze-out is indeed secluded. The relevant process is $\gamma' \leftrightarrow f\bar f$, with rate $\Gamma'$ in the dark photon rest frame. If we assume that the dark photons are relativistic at DM freeze-out, we require that the thermalization temperature given in eq.\eqref{eq:T_rh} is such that  $T'_{\rm rh} \lesssim T'_{\rm fo} = m_{\rm dm}/x'_{\rm fo}$. For $\Gamma' \sim \varepsilon^2 m'$, this gives 
\begin{equation}
\varepsilon\,  m' \lesssim 30\left({m_{\rm dm}^3\over  m_{\rm Pl} x'^{3}_{\rm fo}}\right)^{1/2} \quad \mbox{\rm (no } \gamma'~ \rm{ thermalization)}
 \label{eq:noDPtherm}
\end{equation}
which depends both on $m'$ and $m_{\rm dm}$. As in \eqref{eq:notherm}, the dependence on the DM mass can be traded for the DM abundance, but that would leave a dependence on the HS fine structure constant.  Furthermore, we impose that $m' \lesssim m_{\rm dm}/x'_{\rm fo}$, so that the dark photons are still essentially relativistic at the moment of DM freeze-out. 
The bound from non-thermalization of dark photons is in purple in fig.\ref{fig:constraints}, combining constraint \eqref{eq:noDPtherm} with $m' \lesssim m_{\rm dm}/x'_{\rm fo} \approx m_{\rm dm}/20$.

\subsection{Relevant experimental and astrophysical constraints}
\label{sec:DP_dom_exp}

For fixed $m'$, a lower bound on $\varepsilon$ comes from BBN {constraints}. The dark photons must be non-relativistic before BBN to avoid contributing too much to the expansion rate. This is often expressed in terms of the effective number of neutrino-like degrees of freedom, $\Delta N_{\rm eff}$. Here, we find it more directly relevant to impose that  $\rho'/\rho_{\rm sm} < 0.04$ at $95\%$ confidence at BBN \cite{Hufnagel:2017dgo,Hufnagel:2020nxa}. 

We noted in section \ref{sec:rel_eq} that thermalization before BBN is a necessary but not sufficient condition if it occurs while the dark photons are still relativistic. After thermalization, the dark photons remain in thermal equilibrium with the VS. If they are still relativistic during BBN, then $\rho'/\rho_{\rm SM} = g'_\ast/g_{\ast} \approx 0.3$, which is excluded. Thus, they must be non-relativistic. Requiring $\rho'/\rho_{\rm sm} < 0.04$ sets the lower bound $m'\approx  5$ MeV, corresponding to the shaded vertical line in fig.~\ref{fig:constraints}. If, instead, the dark photons become non-relativistic during the heating of the VS, then the constraint from BBN is essentially that dark photons decay before BBN. Here, $\rho'/\rho_{\rm sm} < 0.04$ so long as $\tau' < 0.2$ sec, so the blue region is excluded in fig.~\ref{fig:constraints}, with a boundary such that $\varepsilon \propto m'^{-1/2}$. The dip around $\sim 90$ GeV is due to the degeneracy between the $\gamma'$ and the Z boson. Extra features correspond to hadronic bound states. To compute the decay rate of the $\gamma'$ into hadrons, we have followed section IV of \cite{Berger:2016vxi} using the R-ratio, $\Gamma(\gamma'\rightarrow \mbox{\rm hadrons}) = R(E_{\rm cm} = m') \Gamma(\gamma' \rightarrow \mu \bar \mu)$. For reference, we also show the bound from a shorter lifetime, $\tau'= 0.02$ s, with the dashed blue line. For this lifetime, thermalization of the dark photons at BBN is only a secondary issue as the expansion is essentially dominated by the SM. It may happen that the dark photons end up thermalizing with the SM if inverse decay processes are significant, see section \ref{sec:nonrel} and appendix \ref{app:FI}. Regardless, after decay, $\rho' \ll \rho$ and the dark photons are totally subdominant.

Finally, we show the  terrestrial collider,  beam dump experiments and astrophysical observations that set relevant bounds on dark photon mixing and mass (see \cite{Cline:2024qzv} for a recent review).  Beam dump experiments provide the most sensitive current limits, see E137 \cite{Bjorken:1988as}, LSND \cite{LSND:1997vqj}, CHARM \cite{CHARM:1985nku}, NuCal \cite{Blumlein:1990ay,Blumlein:2013cua} and \cite{Bauer:2018onh} for a summary. 
These are shown in brown in fig.\ref{fig:constraints}, while the expected improved bounds from SHiP \cite{Alekhin:2015byh} are given by the green dotted line. In turn, the most relevant astrophysical bound comes from the measurement of the timescale of the neutrino burst from SN1987A. 
The corresponding limit we show in yellow in fig.\ref{fig:constraints} is taken from an analysis by Ref.\cite{Hardy:2016kme}. The light green region is from the analysis of \cite{Berger:2016vxi}, which considered that the $\gamma'$ is produced through freeze-in so their abundance satisfies the bound $\rho'/\rho_{\rm sm} < 0.04$. Their analysis used a more sophisticated criteria to constraint dark photons decay around BBN. 

Combining all the constraints discussed in this section, the available mass and mixing parameter space for a $\gamma'$ assuming an initially hot HS, i.e. $T' \gg T$,  corresponds to the white region of fig.\ref{fig:constraints}. 

\subsection{Effects of entropy production}

The entropy produced from a hot HS can be very substantial. In fig.\ref{fig:constraints}, the upper thick black dashed line (R-NR boundary) corresponds to a thermalization temperature such that $3 T'_{\rm th} = m'$, see eq.\eqref{eq:Tth}. This is a convenient criteria to delineate relativistic and non-relativistic thermalization temperatures (again assuming MB statistics). From $3 T'_{\rm th} = m'$ and eq.\eqref{eq:Tth}, we see that the boundary scales  as $\varepsilon \propto m'^{1/2}$, except for $\gamma'$ masses for which it mixes with the $Z$ or hadronic resonances in the photon channel. This line crosses the constraint from DM thermalization, $\varepsilon \lesssim 10^{-6}$, around $m' = 1$ TeV, and the one from BBN around $m' = 5$ MeV and $\varepsilon \sim 10^{-9}$. To the left of that line, thermalization occurs when the dark photons are relativistic (R). For fixed $m'$, the larger the $\varepsilon$, the larger the reheating temperature, $T_{\rm rh} \propto \varepsilon$. Conversely, for fixed $\varepsilon$, the larger the dark photons mass, the larger $T_{\rm rh} \propto m^{2/3}$ (see eq.(\ref{eq:T_rh})). The maximal $T_{\rm rh}$ is reached around $1$ TeV. The lower end corresponds to a HS and VS that thermalize just before BBN, with $m' \sim$ few MeV. 

The separation between the relativistic and non-relativistic scenarios of $\gamma'$ decay is not clear cut. Another possible criteria to separate the two regimes is entropy production. $S_{t,f}/S_{t,i} \approx (g_\ast/g'_\ast)^{1/4}$ in the case of relativistic thermalization, while $S_{t,f}/S_{t,i} \approx (g_\ast/g'_\ast)^{1/4}(\tau'/t_{\rm nr})$ in the non-relativistic regime, where $t_{\rm nr} \sim m_{\rm pl}/m'^2$ is the time when the dark photons become non-relativistic, see sections \ref{sec:entropy_rel} and \ref{sec:entropy_nonrel} and appendix \ref{app:entropyProduction}. In fig.\ref{fig:constraints}, the lower thick black dashed line is set by requiring that $S_{t,f}/S_{t,i}$ departs from $(g_\ast/g'_\ast)^{1/4}$. The region below it corresponds thus to non-relativistic decays, with significant entropy production. The fact that the lower dashed curve does not scale like the upper dashed one is due to changes in $g_\ast$. We call the region between the two thick dashed lines semi-relativistic (SR), a part of the parameter space that could not be covered by our approximate analytical solutions. Contours of constant $S_{t,f}/S_{t,i}$ are shown as thinner dashed lines. Thus, for a large part of the allowed dark photon domain, the entropy dilution factor can be important. This has an impact not only on the DM abundance, but also on any relics which may lie within the VS, such as a baryon asymmetry (see section \ref{sec:baryo}).

To conclude this section, it may be of interest to emphasize that current and future fixed target experiments, notably SHiP, probe  the dark photon parameter range corresponding to relativistic reheating  (R region in the figure). Non-relativistic reheating scenarios (NR region) are much more challenging, as they correspond to dark photons that are both heavy and feebly coupled, and so which stand beyond the current high energy/high intensity frontiers \cite{Lagouri:2022ier}.

\section{Implications for the domain of thermal DM candidates}
\label{sec:domain}

We  now  examine the impact  of heating of the VS on the domain of thermal DM candidates \cite{Coy:2021ann}. 
To recap, the domain delineates in the plane  $\xi_i-m_{\rm dm}$ all possible DM candidates that were in thermal equilibrium in a HS at the moment of their freeze-out, with a temperature ratio $\xi_i$. Its construction is based on generic assumptions, such as the unitarity bound on the DM annihilation cross section. Among other things, it was assumed in \cite{Coy:2021ann} that the Universe was RD from DM freeze-out to BBN. In the language of the present work, this means that the DM companion particles were still relativistic at reheating, see sections \ref{sec:rel}-\ref{sec:constraints} and fig.\ref{fig:constraints}. In that case, entropy dilution is  $S_{t,f}/S_{t,i} \approx (g_\ast/g'_\ast)^{1/4}= {\cal O}(1)$ \cite{Coy:2021ann}. One of our motivations was to check to what extent this is possible. As fig.\ref{fig:constraints} shows, a large part of the $\gamma'$ parameter space corresponds to relativistic reheating.

Heavier and more feebly coupled dark photons leads to non-relativistic reheating, see section \ref{sec:nonrel}. In that case, entropy production can be very important, causing dilution both of possible relics from the VS and of DM itself.  Provided this can be made consistent with a baryogenesis mechanism, such entropy production greatly expands the boundaries of the domain to larger DM masses, both for the case of a hot HS, $\xi \gtrsim 1$, and for a cold HS, $\xi \lesssim 1$.

\begin{figure}[t!]
    \centering
    \includegraphics[width=0.99\columnwidth]{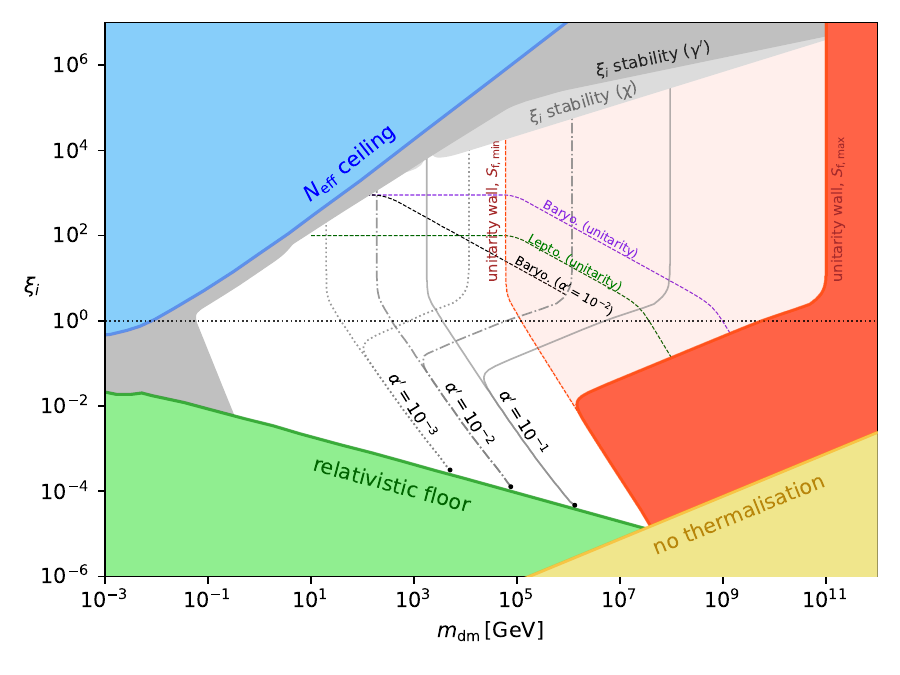}
    \caption{The domain of thermal dark matter candidates, taking into account reheating of the VS. The relativistic floor, $N_{\rm eff}$ ceiling and no thermalization bounds have already been discussed in \cite{Coy:2021ann}, and are not impacted. 
    The unitarity wall (red), which accounts for maximum possible dilution, is shifted to the right with respect to the case of minimum dilution (dashed red line). 
    The same limits in non-unitary cases (i.e. dark QED), for $\alpha'$ = $10^{-1}$, $10^{-2}$ and $10^{-3}$ are shown by the three grey lines (solid, dot-dashed and dotted). As for the unitarity wall, the different curves branch off when entropy production can become significant. 
    The grey $\xi_i$ stability conditions ensure that the initial conditions are stable at the beginning of the scenario.
    The constraints from baryogenesis and leptogenesis (see section \ref{sec:baryo}) are displayed respectively in dashed purple and green for the case of unitary cross section (see also the constraint from baryogenesis for $\alpha'=10^{-2}$ in black).}
    \label{fig:newDomain}
\end{figure}

A second generalisation of the results of \cite{Coy:2021ann} regards the treatment of the temperature ratio at DM freeze-out, $\xi_i$. In \cite{Coy:2021ann}, this was treated as a free parameter and it was assumed that $\xi \approx \xi_i$ is well defined or, in terms of the present work, that the HS and VS were effectively decoupled around DM freeze-out. As we have seen in \ref{sec:plateau}, this occurs so long as the initial heating parameter satisfies $\kappa_i \sim (\rho'_i/\rho_i) \langle \Gamma_i'\rangle/H_i \lesssim 1$. If this is not the case, the HS heats up the VS to a maximal temperature, $T_{\rm max}$, while $\xi$ falls very rapidly. As we shall see, imposing that the HS and VS were decoupled at DM freeze-out sets an new bound on both the maximal and minimal values of $\xi_i$ as a function of the DM mass. That maximal bound is stronger than the one set in \cite{Coy:2021ann} using the constraint from $\Delta N_{\rm eff}$ at BBN. These results, which are detailed in this section, are summarised in fig.\ref{fig:newDomain}, extending the analysis of \cite{Coy:2021ann}. It also includes features that are relevant if baryogenesis takes place before reheating and depicts various dark QED DM candidates, for fixed values of $\alpha'$.

\subsection{Unitarity walls}
\label{sec:UW_S_dilution}

A DM particle that is non-relativistic at freeze-out has a relic abundance that scales as $\Omega_{\rm dm} \propto 1/\langle \sigma v\rangle$.  Assuming a $2$-to-$2$ process, its  annihilation cross-section is  bounded from above by unitarity \cite{Griest:1989wd} 
\begin{equation}
\sigma v  \leq {4 \pi (2 J + 1) \over m^2_{\rm dm} v}.
\label{eq:unitarity_bound}
\end{equation}
Therefore $\Omega \propto m^2_{\rm dm}$, and this sets an upper bound on the  DM mass,  $m_{\rm dm} \lesssim 150$ TeV for s-wave ($J=0$) and assuming SM degrees of freedom only. This is the unitarity bound, it provides a simple and useful way to delineate the possible thermal DM candidates. It is important to keep in mind that higher DM masses are possible, for instance when several partial waves beyond s-wave are relevant, a generic feature for multi-TeV DM candidates \cite{Baldes:2017gzw,Flores:2024sfy}. Here, as in \cite{Coy:2021ann}, we impose the bound \eqref{eq:unitarity_bound} assuming s-wave for simplicity and focus on the impacts of $\xi$ and of possible entropy dilution. 

That the unitarity bound may depend on $\xi_i$ is fairly intuitive. 
If  $\xi_i < 1$, the DM particles are relatively less abundant than for $\xi_i = 1$ and so can be more massive \cite{Hambye:2020lvy,Coy:2021ann}. Explicitly, this bound on $\xi_{i} = \xi_{\rm fo}$ scales like $1/m_{\rm dm}^2$ in the $\xi_i-m_{\rm dm}$ plane, see dotted red curve in fig.\ref{fig:newDomain}. For fixed $\xi_{i}$, a DM particle that is heavier than the unitarity bound is overabundant and thus in principle excluded (the `unitarity wall'). The unitarity bound extends to the region $\xi_{i} \gtrsim 1$. When there is negligible entropy production, the VS plays little role and so the bound is independent of $\xi_{i}$. Soon we will turn to the impact of entropy dilution. 

Using the instantaneous freeze-out approximation, $n_{\chi,{\rm fo}} \langle \sigma v \rangle \approx H_{\rm fo}$, and taking the s-wave unitarity limit for the annihilation cross section, the DM relic density  can be expressed as
\begin{equation}
    \Omega_{\rm dm} h^2 \approx 0.12\, x_{\rm fo}^{\prime 1/2}\,  \frac{({g_\ast/\xi_{i}^4 + g'_\ast})^{1\over 2}}{g_{\ast, s}/\xi_{i}^3 + g'_{\ast,s}}\frac{S_{t,i}}{S_{t,f}} \left(\frac{m_{\rm dm}}{100\,{\rm TeV}}\right)^2
    \label{eq:relic_density}
    \end{equation} 
for all $\xi_i$. The factor that involves the degeneracy parameters, $g_\ast$ and $g'_\ast$, comes from the expansion rate and entropy density, both taken at freeze-out $\xi_i$. 
We also used $\langle 1/v\rangle = x_{\rm fo}^{\prime 1/2}/\sqrt{\pi}$ from MB statistics \cite{Griest:1989wd,Coy:2021ann}. 
Finally,  $S_{t,i}$ ($S_{t,f}$) refers to the total comoving entropy at the initial time (resp. after companion decay), meaning around DM freeze-out.   The relation shows that $\Omega_{\rm dm}$ is independent of $\xi_{i}$ for $\xi_{i} \gg 1$, while $\Omega_{\rm dm} h^2  \sim \xi_{i} m^2_{\rm dm}$ for $\xi_{i} \ll 1$.

\subsubsection{Unitarity wall - no entropy dilution}
If there is no entropy production, setting $g_\ast = 10^2$, $g'_\ast = 3$ and $x'_{\rm fo} = 20$, the unitary limit is 
\begin{equation}
    m_{\rm dm} \approx 60 \, \rm TeV 
\end{equation}
for $\xi_i \gtrsim 1$ and increases toward larger DM masses for $\xi_i \lesssim 1$,
 \begin{equation}
     \xi_i \approx \left( \dfrac{100 \rm TeV}{m_{\rm dm}} \right)^2  
     \label{eq:UW_subdom_entropyless}
 \end{equation}
see fig.\ref{fig:newDomain} and \cite{Coy:2021ann}. For fixed $\xi_{i}$, $x'_{\rm fo} = m_{\rm dm}/T'_{\rm fo}$ is  ${\cal O}(20)$, characteristic of non-relativistic DM freeze-out. However, it can be shown that the DM are less non-relativistic at freeze-out  as $\xi_i$ decreases \cite{Coy:2021ann}.

Another significant bound in the domain is the \emph{relativistic floor} \cite{Hambye:2020lvy}, the green region of fig.\ref{fig:newDomain}. This corresponds to DM particles that freeze-out when they were relativistic, like the SM neutrinos. In that case, $\Omega_{\rm dm} \sim \xi_{i}^3 m_{\rm dm} $ and so $\xi_{i} \propto 1/m_{\rm dm}^{1/3}$. Candidates below the relativistic floor are in principle under-abundant and are therefore excluded, hence the name of the bound. Where the unitarity wall and the relativistic floor meet, the hypothesis that the DM particles were initially in thermal equilibrium breaks down and so that region (in yellow) is also excluded \cite{Hambye:2020lvy,Coy:2021ann}. 

We now consider the impact of possible entropy dilution.  While this depends on the properties of the DM companion, it is possible to draw some generic conclusions, building on what we have learned for the case of a dark photon. 

\subsubsection{Unitarity wall - with entropy dilution and $\xi_i\gtrsim 1$}

Let us begin with $\xi_i \gtrsim 1$ and consider a companion particle that is still relativistic at VS reheating. In this case, the entropy produced only depends on the ratio of the numbers of degrees of freedom of the VS and the DM companion particle,  $S_{t,f}/S_{t,i} \approx (g_\ast/g'_\ast)^{1/4}$. For a massive dark photon,  $g'_\ast = 3$, so for $g_\ast \approx 10^2$ we have $S_{t,f}/S_{t,i} \approx 2.4$. Here, the HS and VS thermalize and the DM companion remains in thermal equilibrium with the SM after reheating. The constraint from BBN requires that the companions are heavier than a few MeV \cite{Hufnagel:2020nxa}. If they meet these conditions, the DM candidate is viable. The unitarity bound on the DM mass is shifted toward a larger maximal DM mass by a factor of $(g_\ast/g'_\ast)^{1/8} \sim {\cal O}(1)$ \cite{Berlin:2016gtr,Coy:2021ann}. 

Consider now $\xi_i \gtrsim 1$, but with a DM companion that is non-relativistic at VS reheating. In that case, significant entropy can be produced. If there is one DM companion, the entropy produced through its decay depends on its lifetime, $\tau'$, and the time $t_{\rm nr}$ when it becomes non-relativistic, see eq.\eqref{eq:entropyHHS}:
\begin{equation}
    \frac{S_{t, f}}{S_{t, i}} \simeq \left( \frac{g_\ast}{g'_\ast} \right)^{1/4}
    \left( \frac{\tau'}{t_{\rm nr}} \right)^{1/2} \qquad (\xi_i \gtrsim 1)
    \label{eq:Entropy1}
\end{equation}
This  reduces to $S_{t, f}/{S_{t, i}} \simeq (g_\ast/g'_\ast)^{1/4}$ if, as in the previous paragraph, reheating occurs when the companion is still relativistic, $t_{\rm nr} \gtrsim  \tau'$. Setting $g'_\ast = 3$ and $g_\ast = 10^2 $ for simplicity, this is rewritten as 
\begin{equation}
    {S_{t,f}\over S_{t,i}} \approx 4 \,\left({m'\over \rm 10\, \mbox{\rm MeV}}\right) \left({\tau'\over 0.2\, \mbox{\rm s}}\right)^{1/2}\qquad (\xi_i \gtrsim 1)
    \label{eq:entropy_hot}
\end{equation}
with $\tau' \lesssim 0.2$ s, a benchmark value that corresponds roughly to BBN time \cite{Hufnagel:2017dgo}. The dependence on $m'$ arises from $t_{\rm nr} \sim 1/H \propto 1/m'^2$ for $\xi_{\rm nr} \gg 1$. Assuming the unitarity limit, the relic abundance reads
\begin{equation}
    \Omega_{\rm dm} h^2 \approx 0.12 \left( \frac{m_{\rm dm}}{100 {\rm TeV}} \right)^2 \left(\frac{10\, {\rm MeV}}{m'} \right) \left({\frac{0.2\,\mbox{\rm s}}{\tau'}}\right)^{1/2}.
    \label{eq:abundanceunitarity1}
\end{equation}
Maximum entropy is produced when the companion is as heavy as possible and has the longest lifetime compatible with BBN. For the companion mass, we take $m'~\sim~T'_{\rm fo} \sim m_{\rm dm}/20$, so that it becomes non-relativistic right at DM freeze-out. The unitarity bound is then
\begin{equation}
    m_{\rm dm} \lesssim\, 10^{11}  \left(\frac{\tau'}{0.2\, \mbox{\rm s}}\right)^{1\over 2}\!\!\text{GeV}  \quad (\xi_i\gtrsim 1,\mbox{unitarity})   \label{eq:newUwall}
\end{equation}
Taking $\tau' \approx 0.2$ s gives the vertical part of the red region in fig.~\ref{fig:newDomain} (maximal entropy production). Comparing with a scenario in which the companion is relativistic at reheating (minimal entropy production), we conclude that the light red region encompasses all DM candidates with  unitarity limit annihilation cross section and a companion lifetime that varies from $\tau' \sim 0.2$ s (maximal entropy dilution) to $\tau'\lesssim t_{\rm nr}$ (minimal entropy dilution). Note that the DM can be much more massive than is usually assumed based on unitarity \cite{Berlin:2016gtr,Bernal:2023ura}

This concerns DM with unitarity limit annihilation cross section (which we called unitary candidates). In the same figure \ref{fig:newDomain}, we give contours of constant $\Omega_{\rm dm} h^2 \approx 0.12$ for dark QED DM candidates with various values of the dark fine structure constant $\alpha'$. As for the unitary candidates, the vertical parts of the curves correspond to minimal, $S_{t,\rm min}$, and maximal, $S_{t,\rm max}$, entropy production. The latter corresponds to dark photons that become non-relativistic around DM freeze-out, $m' \sim m_{\rm dm}/x'_{\rm fo}$ and have maximal lifetime, $\tau' \sim 0.2$ s. The former corresponds to dark photons that thermalize with the VS when they are still relativistic. Between each of these lines, there is a continuum of DM candidates, corresponding to different choices of dark photon mass and decay lifetime (or kinetic mixing parameter).

\subsubsection{Unitarity wall - with entropy dilution and $\xi_i \lesssim 1$}
Consider next $\xi_i \lesssim 1$, meaning a cold HS. Although the main part of our work is focused on a hot HS ($\xi_i \gg 1$), we want  to understand how this region interpolates with the unitarity wall of eq.(\ref{eq:newUwall}). If $\xi_i \ll 1$, we would expect that the companion plays a minimal role, unless it is stable or very long-lived or, alternatively, it is unstable but comes to dominate the expansion of the Universe before its decay, thus producing entropy. We start with the former and consider a light but stable DM companion. Putting aside the possibility that the companion has interactions and could undergo a phase of cannibalism \cite{Hufnagel:2022aiz}, its relic abundance $\Omega'$ is akin to that of DM particles that decouple when they are still relativistic \cite{Hambye:2020lvy,Coy:2021ann}
\begin{equation}
   \Omega' h^2 \approx 0.12\, g'\frac{g_{*s,0}}{g_{*s,\rm fo}}\, \xi_{i}^3\,\left(\frac{ m'}{6 \,\mbox{\rm eV}}\right)
\end{equation}
The factor ${g_{*s,0}}/{g_{*s,\rm fo}}$ takes into account entropy dilution from within the VS, but essentially $\Omega' \sim m' \xi_{i}^3$. As the relativistic floor corresponds to DM candidates with $\Omega_{\rm dm} h^2 \approx 0.12$, fixing $\xi_i$, we conclude that a stable or very long lived dark photon is viable if its mass lies to the left of the relativistic floor, in such a way that $\Omega' h^2 \lesssim 0.12$.

Consider finally $\xi_i \lesssim 1$ with an unstable DM companion that becomes non-relativistic at time $t_{\rm nr}$ and comes to dominate the expansion at time $t_{\rm eq}$. In that case, as explained in appendix \ref{app:entropyProduction}, the entropy produced is given by 
\begin{equation}
    \frac{S_{t, f}}{S_{t, i}} \approx \frac{\rho'_{\rm nr}}{\rho_{\rm nr}} \left( \frac{\tau'}{t_{\rm nr}} \right)^{1/2} 
\end{equation}
see eq.\eqref{eq:entropy_2}. Note that this expression, which is valid provided ${S_{t, f}}/{S_{t, i}} \gtrsim 1$, differs from eq.\eqref{eq:Entropy1} by a factor {$\sim \xi_{\rm nr}^4$} which is much smaller than 1 for $\xi_i \ll 1$. Yet, entropy production can be large if the companion is long-lived. As above, using the  benchmark values $g' =3$ and $g = 10^2$, we express this as
\begin{equation}
    \frac{S_{t, f}}{S_{t, i}} \approx  0.1 \,\xi_i^3 \left( \frac{m'}{10\, {\rm MeV}} \right) \left( \frac{\tau'}{0.2\,{\rm s}} \right)^{1/2}
\end{equation}
This is to be compared with eq.\eqref{eq:entropy_hot}, from which we see that entropy production requires heavier companion mass if $\xi_i \ll 1$.\footnote{\label{foot:FI} This assumes that the abundance of dark companions is fully characterised by $\xi_i$. In other words, we assume that their abundance is not affected by production of companions from VS particles through freeze-in processes. The case of dark photons is instructive, see appendix \ref{app:FI}. As is well-know, abundances through freeze-in processes are controlled by $Y' \sim \Gamma'/H(m') \ll 1$. At the same time, entropy production is maximized when the companions are long-lived, and scales as $(\tau'/t_{\rm nr})^{1/2} \sim ({H(m')/\Gamma'})^{1/2}$. These two effects being antagonist, the entropy produced from particles that are produced through freeze-in is at most $\mathcal{O}(1)$. In other words, we can safely assume that, when entropy production may be substantial,  the abundance of dark companions is fully determined by the initial condition $\xi_i$. We have also checked this numerically.} Consequently, for $\xi_i \ll 1$, assuming unitarity annihilation cross section and entropy dilution, the DM abundance reads
\begin{equation}
    \Omega_{\rm dm} h^2\! \approx\! {0.12 \over\xi_i^{2}} \left( \frac{m_{\rm dm}}{100 \,\rm TeV} \right)^2 \left( \frac{10\, {\rm MeV}}{m'} \right)
    \left(\frac{0.2\,{\rm s}}{\tau'} \right)^{1/2}
\end{equation}
This expression, which applies for $\xi_i \ll 1$, is to be compared with eq.\eqref{eq:abundanceunitarity1}, which is valid for $\xi_i\gg 1$. It contains an extra factor of $1/\xi_i^2$, which stems from eq.(\ref{eq:relic_density}) taking into account entropy production. Both expressions match for $\xi_i \sim 1$.  As for the case where $\xi_i \gg 1$, we see that entropy dilution is maximal if both the companion mass and its lifetime are as large as possible. Taking, as before, $m' \sim m_{\rm dm}/x'_{\rm fo}$ and $\tau' \approx 0.2$ s (maximal entropy production) gives
the part of the  unitarity wall that scales as  
\begin{equation}
    \xi_i \approx \left( \frac{m_{\rm dm}}{10 \,\rm EeV}\right) ^{1/2}\quad ( \mbox{\rm max. entropy dilution})
\end{equation}
see fig.\ref{fig:newDomain}. The maximum allowed DM mass $m_{\rm dm}$ for $\xi_i = 1$ is about $6 \times 10^9 \rm GeV$, similar to values quoted in \cite{Bernal:2023ura}. We found that entropy dilution may become relevant for $\xi_i \approx 10^{-2}$ and $m_{\rm dm} \approx 10^6$ GeV, where the red curves separate in two branches, one corresponding to miminal and the other one to maximal entropy dilution. For smaller values of $\xi_i$, entropy dilution is always negligible in the sense that the companion abundance is so low that it never comes to dominate the expansion of the Universe and so cannot produce significant entropy. {In that case the unitarity wall is as given  by eq.\eqref{eq:UW_subdom_entropyless}.}

Again, we have focused on the impact of entropy dilution on the unitarity wall, but clearly the same features arise for specific DM candidates. Fig.\ref{fig:newDomain} shows contours of constant $\Omega_{\rm dm} h^2\approx 0.12$ curves for different cases of dark QED. We see the similar branching between DM candidates with little entropy production ($S_{\rm t, min}$), and those with maximal entropy production, ($S_{\rm t, max}$), with a continuum of candidates between the two branches. We also observe that when the fine structure constant decreases and the curves move toward the left part of the domain, the gap between the branches shrinks. This is because the maximal amount of entropy that can be produced decreases as the maximal dark photon mass decreases.

\subsection{Stability of the initial temperature ratio}
\label{sec:stability}

One of the  basic assumptions underlying the domain depicted in fig.\ref{fig:newDomain} is that the initial temperature ratio between the HS and VS is a well-defined parameter. Concretely, we will require that at early times (i.e. around the time of DM freeze-out) the temperature ratio does not evolve rapidly, $\xi\approx \xi_i$. We have called this the plateau in previous sections. In section \ref{sec:attractor}, we introduced the `heating parameter' $\kappa$, see eq.(\ref{eq:kappa_def}). We showed that this simple parameter controls the onset of heat transfer from the HS to the VS. If initially $\kappa_i \lesssim 1$, then there is a plateau, $\xi \approx \xi_i$. If instead $\kappa_i\gtrsim 1$, the energy transfer from the HS to the VS is efficient and the temperature of the VS ($\xi$) rapidly increases (resp. decreases) to some maximal temperature, superseding the initial VS temperature $T_i$, see purple curve in fig.\ref{fig:relativistic}. This is for $\xi_i \gg 1$ but a similar condition holds for $\xi_i \ll 1$, the difference being that the role of the VS and HS are reversed. We begin by considering the case $\xi_i \gg 1$.

We now use the heating parameter to impose an extra condition on the domain by requiring that the initial temperature ratio is stable. For $\xi_i \gg 1$, we demand
\begin{equation}
    \kappa_i \approx \frac{\rho'_i}{3 \rho_i} \frac{\langle\Gamma'\rangle_i}{H_i} \lesssim 1
\end{equation}
at $T'_i \approx m_{\rm dm}/x'_{\rm fo}$. This corresponds to
\begin{equation}
    0.05\,\xi_i^4\, \left( \frac{m'}{10\,{\rm MeV}} \right) \left( \frac{\rm GeV}{m_{\rm dm}} \right)^3 \left( \frac{0.2 \,\rm s}{\tau'} \right) \lesssim 1.
    \label{eq:stability1}
\end{equation}
Again, we set $g_\ast = 10^2$, $g'_\ast = 3$ and $x'_{\rm fo}\equiv x'_i = 20$ to avoid cluttering; the factor $m'/m_{\rm dm}$ is due to  time dilation.  This condition depends on $m'$ and $\tau'$. These parameters can be eliminated by noting that the condition is the weakest, thus leaving the largest domain, provided $m'$ is as light as possible and $\tau'$ as long as possible. As we have seen, we must require that the companion decays and is non-relativistic  before BBN. Taking $m' \approx 5$ MeV and $\tau' =0.2$ s and assuming no entropy production gives  
\begin{equation}
    \xi_i \lesssim  \left( \frac{m_{\rm dm}}{100\,\rm MeV} \right)^{3/4} \quad (\xi_i \gtrsim 1, \mbox{\rm no entropy dilution})
\end{equation}
This condition corresponds to the grey region depicted in fig.\ref{fig:newDomain} for {$m_{\rm dm} \lesssim 100\, \rm TeV$ and $\xi_i >1$. 

Heavier DM candidates are possible, provided there is entropy dilution. Within the the red-shaded region, we impose stability of $\xi_i$ making  also sure that the companion decay will produce the right amount of entropy. Here the dark sector is largely dominant, so the expression (\ref{eq:abundanceunitarity1}) holds, in which we can isolate $m'$ and inject it into (\ref{eq:stability1}) to obtain
\begin{equation}
    \xi_i \lesssim 10^5 \left( \frac{m_{\rm dm}}{\rm 100\, TeV} \right)^{1/4} \quad (\xi_i \gtrsim 1, \mbox{\rm with entropy dilution})
\end{equation}
The two conditions cross each other at $m_{\rm dm} \sim 100$ TeV, as expected.

The same consideration applies to the DM particles themselves. Concretely, we must also make sure that substantial energy transfer does not happen before DM freeze-out via DM-SM particles scatterings. To do so, similarly to the condition on the heating parameter for dark photon decay, we require that 
\begin{equation}
   \kappa_{\rm dm} \sim \left. \frac{\rho_{\rm dm}}{\rho} \frac{\Gamma}{H} \right|_{T' \sim m_{\rm dm}} \lesssim 1
   \label{eq:kappaDM}
\end{equation}
with $\rho_{\rm dm}/\rho \sim \xi^4$ and here $\Gamma \sim \alpha \alpha' \varepsilon^2 m_{\rm dm}$ is the  interaction rate between DM and SM particles around $T' \sim m_{\rm dm}$. To make this condition as conservative as possible we set $m' = m_{\rm dm}/x'_{\rm fo}$ and $\tau' = 0.2 \rm s$, and determine the value of $\alpha'$ by requiring the right relic abundance for a non-relativistic freeze-out \cite{Kolb:1990vq}:
\begin{equation}
    \Omega_{\rm dm} h^2 \simeq 10^9 \frac{x_{\rm fo}}{{g'_\ast}^{1/2} m_{\rm Pl} \left< \sigma v \right>} \frac{S_{t,i}}{S_{t,f}} = 0.12
\end{equation}
with $\left< \sigma v \right> \sim \alpha'^2/m_{\rm dm}^2$ the annihilation of DM into dark photons. In that case, where the DS is dominant, the entropy production due to dark photons decay follows (\ref{eq:entropyHHS}). Taking the same values for $g_\ast$, $g'_\ast$ and $x'_{\rm fo} \sim 20$ as usual, we obtain the condition\footnote{
This bound is obtained from eq.\eqref{eq:kappaDM}, in which the dependence on the factor $\varepsilon^2$ is eliminated by setting the lifetime of the dark photon to be around BBN (see section \ref{sec:DP_dom_exp} and the boundary of the blue region in figure \ref{fig:constraints})}
\begin{equation}
    \xi_i \lesssim \left( \frac{m_{\rm dm}}{100 \rm eV} \right)^{3/8} \quad (\xi_i \gtrsim 1, \text{with entropy dilution})
\end{equation}
which is depicted by the light grey region in figure \ref{fig:newDomain}, labeled \emph{$\xi_i$ stability ($\chi$)}.

Next, we consider a cold HS, $\xi_i \lesssim 1$. In that scenario, it is possible that the HS is completely secluded (i.e. $\varepsilon \rightarrow 0$) provided the dark photon mass lies within the relativistic floor (see above). If not, the dark photon should decay into SM particles and this without spoiling the abundance of elements produced by BBN. This problem is beyond our scope (but is work in progress). Here we  focus on the stability of the initial temperature ratio. The reasoning of section \ref{sec:attractor} straightforwardly applies, provided the role of the two sectors are inverted, that is $\rho' \leftrightarrow \rho$. This leads to  the following condition for the stability of $\xi_i$
\begin{equation}
    \kappa'_i \approx\frac{\rho_i}{3 \rho'_i} \frac{\langle \Gamma' \rangle}{H_i} \lesssim 1.
    \label{eq:FIcondition}
\end{equation}
In this expression, the relevant rate is driven by the conditions of the VS. Assuming, as for the dark photon, that the relevant process for energy transfer from the VS to the HS is production through inverse decay, we have $\langle\Gamma' \rangle \sim m'\Gamma'/T$. In plain words, the condition of stability is simply that freeze-in production can be neglected at the time of DM freeze-out, $\xi_i \equiv \xi_{\rm fo} \ll 1$. With the same approximations as before on $g_\ast$, $g'_\ast$ and $x'_{\rm fo}$, and the same choice for $m'$ and $\tau'$, this gives
\begin{equation}
    \xi_i \gtrsim \left( \frac{100 \,\rm MeV}{m_{\rm dm}} \right)^3 \qquad (\xi_i \lesssim 1)
    \label{eq:FIcondition2}
\end{equation}
see grey region for $\xi_i \ll 1$ in fig.\ref{fig:newDomain}. For $\xi_i \ll 1$, entropy production is only relevant for much heavier dark companions. As explained above, for such particles, production through freeze-in is negligible, see appendix \ref{app:FI}.\footnote{Even if the initial temperature $\xi_i \ll 1$ is stable at DM freeze-out, there may be subsequent production of dark companions through freeze-in thus superseding the initial abundance of dark companions, see \ref{app:FI}. If the dark companion is unstable, as for the case of the dark photon considered in the work, the very same process that lead to dark companions production will lead to their decay. This comes with no or negligible entropy production and so has no impact on the DM abundance and thus on the domain, see footnote \ref{foot:FI}.} Finally,  we have checked that the freeze-in production of DM leads to a condition of stability that is parametrically weaker than that from the freeze-in of dark photons.

\section{Comments on baryogenesis and leptogenesis}
\label{sec:baryo}

From BBN and CMB anisotropies, the baryon-to-photon ratio of the Universe is \cite{ParticleDataGroup:2020ssz},
\begin{equation}
    \eta \equiv {n_b}/{n_\gamma} \approx 6 \times 10^{-10}
\end{equation}
corresponding to a baryon number $B= (n_b-n_{\bar b})/s \approx 10^{-10}$ in the early Universe, 
a tiny number that could be explained through some baryogenesis mechanism \cite{Kolb:1990vq}. In most scenarios, baryogenesis is supposed to take place in the very early Universe, through the decay of some very heavy particle, like in leptogenesis \cite{Davidson:2008bu}, at the electroweak scale \cite{Morrissey:2012db}, or from some bayon-number charged scalar field (Affleck-Dine) \cite{Dine:2003ax}. However, motivated by a low reheating scenario \cite{Hannestad:1995rs}, there are also baryogenesis mechanisms that try to explain the generation of the baryon asymmetry at  temperatures as low as $\sim$ few MeV, see e.g. \cite{Elor:2018twp} and references therein. Assuming the former class of mechanisms, a hot HS has several  interesting implications for baryogenesis if it takes place before DP decay. 

We can parameterize the baryon number at the moment of baryogenesis as $B = \epsilon_B T_B^3/s_{B}$ where $\epsilon_B$ captures baryon number violation and C and CP violation factors, $T_B$ is the characteristic temperature and $s_{B}$ the entropy density. On general grounds, $\epsilon_B$ is a small number. For instance, if $s_B \sim g_\ast T^3_B$ with $g_\ast \sim 10^2$, then $\epsilon_B \sim 10^{-8}$ or so is required to explain the baryon number. Thus $\epsilon_B$ is essentially a measure of the produced baryon asymmetry \cite{Kolb:1990vq}. Baryogenesis/leptogenesis scenarios that involve the baryon/lepton number decay of some heavy particle lead naturally to such a small $\epsilon_B$, but in scenarios {\em à la} Affleck-Dine, the asymmetry can be much larger, possibly $\epsilon_B \sim 1$. Now, if the expansion is driven by a hot HS, then $s_B \sim g'_\ast T'^3_B$ and so $B \sim \epsilon_B/(g'_\ast \xi_B^3)$. This has obvious implications. 

If thermalization of the VS and HS occurs when the HS particles are relativistic, entropy production is ${\cal O}(1)$ and so the temperature ratio must be such that $\xi_B \approx ( 10^{8} \epsilon_B)^{1/3}$. Imposing that, at most, $\epsilon_B \lesssim 1$ puts an upper bound  on the temperature ratio, $\xi_i \lesssim 10^3$. We report such constraints in figure \ref{fig:newDomain} by assuming that $\xi_B \sim \xi_i$. If thermalization occurs when the HS particles are non-relativistic, entropy production takes in and the baryon asymmetry is further reduced by a factor $(S_{t,i}/S_{t,f})$. We extract this factor by fixing the DM abundance (see the diagonal part of the dashed lines), e.g. using eq.\eqref{eq:relic_density}. In that case (purple dashed), the upper bound on $\xi_i$ scales as $\xi_i \lesssim 10^3 \left(100\, \rm TeV/m_{\rm dm} \right)^{2/3}$ for $\xi_i \gg 1$, and $\xi_i \lesssim (1\, \rm EeV/m_{\rm dm})^2$ for $\xi_i \lesssim 1$, maximum entropy dilution corresponding to the crossing with the boundary of the unitarity wall. From this, we see that baryogenesis puts an upper bounds on the maximal DM mass, which range from $m_{\rm dm} \lesssim 10^5 $ GeV for $\xi_i \sim 10^3$ to $m_{\rm dm} \lesssim 10^9 $ GeV for $\xi \sim 1$. We also depict (black dashed) the case of entropy dilution from a dark photon with a more mundane coupling, here $\alpha' = 10^{-2}$. There, the maximum entropy dilution corresponds to the crossing with the branch of the dot-dashed curve with maximal entropy dilution, see figure \ref{fig:newDomain}. 

A hot HS may also impact the dynamics of baryogenesis/leptogenesis. For instance, the condition for out-of-equilibrium decay of some very heavy particle, say $N$, becomes $\Gamma_N \lesssim H(T \sim m_N) \sim g'_\ast \xi^2 m_N^2/m_{\rm Pl}$. Due to the faster expansion rate, this implies that the decaying particle could be substantially lighter than in the case of ordinary expansion. With $\Gamma_N \sim \alpha_N m_N$, the condition for out-of-equilibrium decay becomes $m_N \gtrsim \alpha_N m_{\rm Pl}/\xi^2$. 
Note that if the inflaton decays dominantly into the HS, imposing $T'_{\rm rh} \lesssim 10^{15}$ GeV, see e.g. \cite{Lozanov:2019jxc}, requires only that $T_{\rm rh} \lesssim 10^{15}/\xi$ GeV. There is thus room to accommodate $m_N \lesssim T_{\rm rh}$.

Interestingly, the situation is more contrived in the case of leptogenesis because of the link  to the origin of the mass of SM neutrinos. We discuss only a very basic scenario, assuming the risk of oversimplification. We follow the notations of \cite{Strumia:2006qk}. Consider the decay of a heavy Majorana neutrino $N_1$ of mass $M_1$ into the Higgs and lepton doublets. At tree level $\Gamma_1 \sim \lambda_1^2 M_1$ with $\lambda_1$ the relevant combination of Yukawa couplings. Through the see-saw mechanism, this Yukawa parameter can be expressed in term , $\tilde m_1 = \lambda_1^2 v^2/M_1$, the contribution of $N_1$ to the mass of SM neutrinos. With a hot HS, the ratio of the decay rate to the expansion rate is $R = \Gamma_1/H(M_1) \sim {\tilde m_1/ \tilde m^\ast \xi^2}$ with $\tilde m^\ast \sim v^2/m_{\rm pl} \sim 10^{-3}$ eV. If $R \lesssim 1$, the $N_1$ decays  far from thermal equilibrium; if $R \gtrsim 1$, the lepton asymmetry is suppressed by a factor of $\eta \approx 1/R$, with $\eta$ the efficiency factor \cite{Strumia:2006qk}.  As concluded in the previous paragraph, the faster expansion rate makes the condition $R \lesssim 1$ easier to satisfy. In the parlance of \cite{Strumia:2006qk}, efficiency is $\eta = 1$. At the same time (first remark) it is essentially independent of the mass of the $N_1$ state, provided $\tilde m_1$ is characteristic of the mass of SM neutrinos. This is model dependent, but there is also an upper bound on $\tilde m_1 \lesssim 0.1$ eV from measurements of SM neutrino masses. This implies that for $\xi \gtrsim 10$, the decay of $N_1$ is always strongly out-of-equilibrium and the efficiency of leptogenesis is ${\cal O}(1)$. The other relevant parameter is the CP asymmetry from $N_1$ decay, which arises from interference between tree and loop amplitudes, $\epsilon_1 \sim M_1/M_{2,3} \, \mbox{\rm Im}(\lambda^2_{2,3})$ with $M_{2,3} \gg M_1$ the mass of heavier Majorana neutrinos $N_{2,3}$. In terms of the contribution of these heavy $N$ to the mass of SM neutrinos, $\tilde m_{2,3} = \lambda_{2,3}^2 v^2/M_{2,3}$, $\epsilon_1 \sim M_1\, \mbox{\rm Im}(\tilde m_{2,3})/v^2$. As the lepton asymmetry is $\sim \epsilon_1/\xi^2$, we see that a hotter HS requires a larger $\epsilon_1$ and, consequently, tends to require a {\rm heavier} $N_1$ particle (second remark). This is again model dependent, but as shown in  \cite{Davidson:2002qv}, there is an upper bound on the CP parameter, $\epsilon_1 \leq 3 M_1 (m_{\nu_3} - m_{\nu_1})/16 \pi v^2$ where $m_{\nu_{3,1}}$ are the mass of the heavier and lighter SM neutrinos. This translates into a lower bound on the mass of $N_1$, $M_1 \gtrsim 10^9 \xi^2$ GeV, which is the Davidson-Ibarra bound applied to a hot HS. From $M_1 \lesssim T_{\rm rh} \lesssim 10^{15}/\xi$ GeV, we get that the leptogenesis mechanism described here requires that $\xi \lesssim 10^2$. This assumes that there is not further entropy dilution from HS particles decay. If so, the effect is as discussed above. The limit from leptogenesis is reported as  green dashed line in figure \ref{fig:newDomain} in which we assume unitarity limit on the lifetime of the HS companion. 

In a different class of scenarios, baryon-number violating processes around the electroweak phase transition are also affected if the expansion of the Universe is non-standard \cite{Davidson:2000dw}. As mentioned above, the temperature for relativistic thermalization may lie somewhere between a few MeV and around $1$ TeV. That range includes the scale for EW breaking,  $T_c \sim 100$ GeV \cite{Kajantie:1996mn}, so it may have occurred while the expansion was dominated by a hot HS, impacting scenarios for EW baryogenesis. In particular, the rate for sphaleron transitions in the unbroken phase is $\Gamma_s \sim \alpha_W^5 T$ \cite{Arnold:1996dy} while $H \sim g_\ast'^{1/2} T'^2/m_{\rm Pl}$, so if there is a  hot HS, they are efficient at lower temperatures, $T \lesssim 10^{10} \rm{GeV}/\xi^2$. In the broken phase, their rate is Boltzmann suppressed, $\Gamma_s \propto \alpha_W^5 T e^{-E_s/T}$ with $E_s \sim m_W/\alpha_W$. In scenarios with a first order EW phase transition, sphalerons are typically required to be out-of-equilibrium inside bubbles of true vacuum \cite{Morrissey:2012db}. This condition can be expressed as a condition on the jump of the vev $v_b$ of the Higgs field at the moment of bubble nucleation, $v_b/T_c  \gtrsim 1 $ \cite{Arnold:1996dy,Davidson:2000dw}. If the expansion is dominated by a hot HS, sphaleron processes are more easily out-of-equilibrium, thus relaxing the bound on the change of the Higgs field. A direct adaptation of the argument in \cite{Davidson:2000dw}, see eq.(3,4), gives $v(T_b)/T_c \gtrsim (0.9,0.8,0.6)$ for $\xi = (10,10^2,10^3)$ respectively. This trend shows that sphaleron processes freeze-out more efficiently in the presence of a hot HS. 

In this section, we simply assumed that the expansion is along a plateau, with $\xi \approx $ constant. This could also occur along the period of heating of the VS, during which $\xi \propto a^{-3/4}$ and the temperature of the VS evolves very slowly, $T \propto a^{-1/4}$. As figs. \ref{fig:evolution} shows, the process of heating may last long, taking typically several scale factor decades, so several VS sector phenomena may occurs along such an era (QCD phase transitions, etc.) provided it happens before BBN.  We leave such considerations for future work. 

\section{Conclusions}

We have studied further the possibility that DM belongs to a hidden sector at a temperature $T'$ that differs from that of the VS or SM, $T$, with a focus on a relatively hot HS scenario, $\xi = T'/T \gg 1$. This setup is by and large motivated by \cite{Coy:2021ann} in which it was assumed that the HS contains a dark matter particle with an abundance that is set by secluded thermal freeze-out through annihilation into a lighter particle, or companion.
  The key issue that we have studied is the fate of the DM companion, and in  particular how its energy is transferred to Standard Model particles, reheating the VS. 

Reheating from particle decay is a rather standard problem in cosmology. Our analysis complements the study of similar DM scenarios in the literature, see in particular \cite{Berlin:2016gtr,Heurtier:2019eou,Bernal:2023ura}. First and foremost, we have systematically studied the different eras in the history of the hot HS, starting from  $\xi_i = T'_i/T_i \gg 1$, down to reheating of the VS and the subsequent thermalization of the DM companion. We did so both numerically and analytically, using for the sake of concreteness the  framework of dark QED coupled to the SM through kinetic mixing. In brief, we have distinguished  two scenarios, which we dubbed relativistic and non-relativistic reheating, depending on whether the companion (dark photon in the framework of QED) was relativistic or non-relativistic at the end of reheating. Along relativistic reheating, the expansion of the Universe is radiation dominated, with $\xi \approx $ constant, but at a much faster rate than in standard cosmological scenarios. Doing so, we  emphasized the relevance of a specific  combination of rates and energy densities that controls the process of energy transfer and which we called the heating parameter $\kappa$, see eq.\eqref{eq:heatpar}. In particular, we have shown that the initial temperature ratio remains constant, $\xi \approx \xi_i$ as long as $\kappa \leq 1$ and that $\kappa \sim 1$ between the onset and the end of heating of the VS. Of course, a similar parameter holds for scenarios in which the VS is hotter than the HS, as discussed in appendix \ref{app:FI}.
Finally, we have articulated these two scenarios to the 'all important problem of entropy production' \cite{Kolb:1990vq}. While the underlying physics is well-know, we have put forward the fact that entropy production scales essentially as $\sqrt{\tau'/\tau_\ast}$, where $\tau'$ is the lifetime of the decaying particle and $\tau_\ast$ is some characteristic time which depends on the problem at hands. For instance, in rather standard scenarios, $\tau_\ast$ corresponds  to the moment when the decaying particle starts to dominate the expansion of the Universe, see appendix \ref{app:entropyProduction} for details. 
Our main results regarding the process of reheating from a HS are illustrated in figures \ref{fig:XiSchematic} and \ref{fig:evolution}. 

Next, we have used our results to delineate the possible parameter space of the dark photons, meaning its mass $m'$ and the kinetic mixing parameter $\varepsilon$, together with astrophysical and laboratory constraints, cf. figure \ref{fig:constraints}. From this, we can read in particular for which  parameters dark photons are consistent with a relatively hot HS scenario and, if so, whether they could lead to relativistic reheating or non-relativistic reheating. Interestingly, current and forthcoming fixed targets experiments, like SHiP, are essentially able to probe scenarios in which the VS went through relativistic reheating. Indeed, non-relativistic reheating corresponding to heavy and very feebly coupled dark photons (more generally, DM companions) are and will remain beyond reach of experiments. 

Our initial goal was to study the implications of VS reheating for the domain of thermal DM candidates, in the parlance of \cite{Coy:2021ann}. 
 One of our main results concerns the highest possible temperature ratio as a function of the DM mass, see figure \ref{fig:newDomain}. In \cite{Coy:2021ann}, a very general bound was set by requesting that the HS is not dominating the expansion of the Universe by the time of BBN, based on constraints on $\Delta N_{\rm eff}$. In the present work, we have studied to what extent a given fixed temperature ratio can be a well-defined initial condition, a stronger and more realistic requirement than the bound on $\Delta N_{\rm eff}$. Next, we have shown that the domain of possible thermal dark matter candidates can extend to very large masses, provided the VS is reheated short before BBN. In particular, they could be as heavy as $\sim 10^{11}$ GeV, which is much heavier than the standard unitarity upper bound  on the WIMP mass \cite{Griest:1989wd}. This is not {\em per se} a new result \cite{Berlin:2016gtr,Heurtier:2019eou,Bernal:2023ura} but the application to the domain of thermal DM particles is new. In particular, in our work we cover and articulate both the cases of $\xi \gg 1$ (hot HS) and $\xi \ll 1$ (cold HS) which both may lead to super heavy DM candidates. Unfortunately, all such DM candidates and their companions are also very feebly coupled to the VS, and consequently, they are way beyond the current energy and intensity frontiers, unless we allow them to decay with a very long lifetime, a scenario that is possible but that we have not considered because it is specific. However, as we have already emphasized, DM companions that could lead to scenarios with relativistic reheating may be within reach of current and forthcoming experiments, meaning SHiP.

Focusing on relativistic reheating, a legitimate question is whether it could have phenomenological implications? The main point of such scenarios is that the expansion of the Universe is  radiation dominated but is much faster than usual, a possibility that has been already considered in the literature for various but specific phenomenological works, see e.g. \cite{Berlin:2016gtr,DEramo:2017gpl,Bernal:2023ura,Bringmann:2023iuz}. Here, we have focused on generic implications, in particular on the DM mass and on the temperature ratio. From that perspective, we discussed the possible implications for simple baryogenesis and leptogenesis scenarios, in particular if the baryon asymmetry is created while the expansion is still dominated by a hot HS, in which case it is possible to set extra bounds on the maximal temperature ratio. In particular, we worked out a generalisation of the Davidson-Ibarra bound set by a phase of fast relativistic expansion. In particular, we show that a leptogenesis mechanism, at least in its simplest incarnations, is possible only provided the temperature ratio was of the order of $\xi_i \lesssim 10^2$, see figure \ref{fig:newDomain}.

\label{sec:prospects}
%



\section*{Acknowledgments}
We thank Thomas Hambye and Marco Hufnagel for discussions and collaboration. This work has been supported by the F.R.S./FNRS under the Excellence of Science (EoS) project No. 30820817-be.h “The H boson gateway to physics beyond the Standard Model”, by the FRIA, and by the IISN convention No. 4.4503.15.


\appendix

\section{Summary of Maxwell-Boltzmann statistics}
\label{app:therm}

In this work, we made use of simplifications offered by Maxwell-Boltzmann (MB) statistics.
First,  all basic thermodynamical densities can be expressed in terms of simple functions \cite{Gondolo:1990dk, Kolb:1990vq}, in which the chemical potential factors out,\footnote{{These quantities  may refer to either HS or VS ones; here we drop the primes that we use in the bulk of the text to refer to HS sector quantities.}} 
\begin{eqnarray}
n &=& \frac{g}{2\pi^2} \frac{m^3 }{x} K_2(x) e^{\mu/T} \label{eq:neq} \\ 
\rho &=& \frac{g}{2\pi^2} m^4 \left[ \frac{1}{x} K_3(x) - \frac{1}{x^2} K_2(x) \right] e^{\mu/T} \label{eq:rhoeq}\\
&\equiv& {g\over 2 \pi^2} m^4\left[ \frac{1}{x} K_1(x)+\frac{3}{x^2} K_2(x) \right] e^{\mu/T}\nonumber\\
p &\equiv& n T =  \frac{g}{2\pi^2} \frac{m^4}{x^2} K_2(x) e^{\mu/T} \\
s &=& {\rho  + p - \mu n\over T} \notag \\
&=& \frac{g}{2\pi^2} \left(m^3  K_3(x) - \mu \, m^2  K_2(x)\right) e^{\mu/T} \, ,
\end{eqnarray}
where the $K_\nu(x)$ are modified Bessel functions and $x = m/T$.
Their asymptotic forms are
$$
K_\nu(x) \sim \left({\pi\over 2 x}\right)^{1/2} e^{-x}\left(1+{4 \nu^2-1\over 8 x} + {\cal O}(1/x^2)\right)$$ for large $x$ and 
$$
K_\nu(x) \sim {1\over2 } \Gamma(\nu) \left({2\over x}\right)^{\nu}$$
for small $x$. 

The equation of state of a specie is given by
\begin{equation}
    w \equiv \frac{p}{\rho} = \frac{\rho_{\rm eq}}{p_{\rm eq}} = \frac{K_2(x)}{x K_3(x) - K_2(x)} \, ,
\end{equation}
which is independent of $\mu$. Also, $p=n T$ holds both in the relativistic and the non-relativistic periods using MB. At low temperatures, $m\ll T$, $\rho = m n$ with $n$ the usual density of non-relativistic particles, 
\begin{equation}
n = g \left({mT\over 2 \pi}\right)^{3/2} e^{- (m-\mu)/T}
\label{eq:neqNR}
\end{equation}
while at large temperatures, $T \gg m$, 
\begin{eqnarray}
n &=& {g \over\pi^2}  T^3 e^{\mu/T}  \\
\rho &=& 3 n T = 3 p \\
s &=& (4 - \mu/T) n
\end{eqnarray}
This is where MB departs from Fermi-Dirac (FD) and Bose-Einstein (BE), since the inter-particle separation is of the order of typical wavelengths, $\lambda \sim n^{-1/3} \sim 1/T$, and quantum statistics effects cannot be neglected. 
However, the error made using MB statistics in this relativistic period is only ${\cal O}(10 \%)$.\footnote{For the sake of comparison, for $\mu =0$, $$ n_{\rm eq}/gT^3 = (0.091, 0.10, 0.12)$$
$$ \rho_{\rm eq}/gT^4 = (0.29,0.30,0.33) $$ 
$$ s_{\rm eq}/gT^3 = (0.38,0.41,0.44) $$ where the entries refer respectively to FD, MB and BE statistics. MB stands in between FD and BE. This is yet another motivation to use MB:  it gives numbers that are midway between  fermionic and bosonic particles.  While we are comparing different statistics, it is interesting to notice that the relative error made in setting $s = g^{1/4} \rho^{3/4}$ is only $(0.04,-0.01,-0.01)$ again for FD, MB and BE statistics. We make use of this in the appendix on entropy production. \label{foot:MB_approx}}

Finally, the mean energy of the particle is given by
\begin{equation}
    {\langle E \rangle\over m} = \frac{\rho}{m n} = \frac{\rho_{\text{eq}}}{m n_{\text{eq}}} = \left\{\begin{array}{c}
   1+{3\over 2 x} = 1  + {3\over 2} {T\over m} \qquad x \gg 1\\
    \\
    {3\over x}= {3 T\over m} \qquad x \ll 1
    \end{array}\right.
\end{equation}
independent of $\mu$. For the thermally averaged decay rate, we also need 
\begin{equation}
\left\langle{1\over \gamma}\right\rangle = \left\langle {m\over E}\right\rangle  = {K_1(x)\over K_2(x)}= \left\{\begin{array}{c}
    1- {3\over 2 x} = 1  - {3\over 2} {T\over m}  \qquad  x \gg 1\\
    \\
    {x\over 2}= {m\over 2 T}\qquad x \ll 1 
    \end{array}\right.
\end{equation}
For these expressions, it is important to keep the $\mathcal{O}(1/x)$ in the asymptotic form of the Bessel functions for large $x =m/T$.

\section{Cross sections}
\label{app:cross_sec}
{In this work, we consider three main processes : dark electron pair annihilation into dark photons (for DM freeze-out, cross section $\sigma_{\chi \overline{\chi} \rightarrow f \overline{f}}$), dark electron pair annihilation into SM fermions (for the $\chi$ thermalization constraint, cross section $\sigma_{\chi \overline{\chi} \rightarrow \gamma' \gamma'}$), and dark photon decay into SM fermions (decay rate $\Gamma'$). 

\subsection{Dark electrons annihilation into dark photons}
The cross section for the annihilation of dark electrons into a pair of massive dark photons is \cite{Aboubrahim:2021ycj}
\begin{align}
    \begin{split}
        \sigma_{\overline{\chi} \chi \rightarrow \gamma' \gamma'} &= \dfrac{4 \pi \alpha'^2}{s} \left\{ \dfrac{s^2 + 4 m_{\rm dm}^2  (s-m'^2) + 4 m'^4 - 8 m_{\rm dm}^4}{(s-2m'^2)(s-4m_{\rm dm}^2)} \ln B \right. \\
        & \left. - \sqrt{\dfrac{s-4m'^2}{s-4m_{\rm dm}^2}} \left[ \dfrac{m_{\rm dm}^2 s + 2 m'^4 + 4 m_{\rm dm}^4}{(s-4m'^2)m_{\rm dm}^2 + m'^4} \right] \right\}
    \end{split}
    \label{eq:sigma_XX_AA_massive}
\end{align}
with
\begin{equation}
    B = \dfrac{s - 2m'^2 + \sqrt{(s-4m'^2)(s-4m_{\rm dm}^2)}}{s - 2m'^2 - \sqrt{(s-4m'^2)(s-4m_{\rm dm}^2)}}
\end{equation}
In the massless dark photon limit, this expression becomes \cite{Chu:2011be}
\begin{equation}
    \sigma_{\chi \overline{\chi} \rightarrow \gamma' \gamma'} = \dfrac{4\pi \alpha'^2}{s} \left[ \dfrac{2 s^2 + 8 m_{\rm dm}^2 s - 16 m_{\rm dm}^4}{s (s - 4 m_{\rm dm}^2)} \tanh^{-1} \left( \sqrt{1 - \dfrac{4 m_{\rm dm}^2}{s}} \right) - \dfrac{s + 4 m_{\rm dm}^2}{\sqrt{s (s - 4 m_{\rm dm}^2)}} \right]
    \label{eq:sigma_XX_AA}
\end{equation}
 The numerical error due to the use of this approximation around freeze-out temperatures, $T' \sim m_{\rm dm}/20$, is less than $\mathcal{O}(0.1 \%)$. In the non-relativistic limit, relevant for the DM freeze-out process, it reduces to \cite{Pospelov:2007mp}
\begin{equation}
    \sigma_{\chi \overline{\chi} \rightarrow \gamma' \gamma'} v \simeq \dfrac{\pi \alpha'^2}{m_{\rm dm}^2} \sqrt{1 - \dfrac{m'^2}{m_{\rm dm}^2}}
\end{equation}

\subsection{Dark photons annihilation into SM fermions}
For the annihilation process of dark photons into SM fermions, the cross section in the massless dark photon limit reads \cite{Chu:2011be}
\small
\begin{align}
    \begin{split}
        \sigma_{\chi \overline{\chi} \rightarrow f \overline{f}} &= \sum_f N_f^C \dfrac{\pi q_e^2 \alpha \alpha' \varepsilon^2}{s} \sqrt{\dfrac{s - 4 m_f^2}{s - 4 m_{\rm dm}^2}} \\
        & \times \left\{ \left(1 + \dfrac{s^2 - 4 m_f^2 s - 4 m_{\rm dm}^2 s + 16 m_f^2 m_{\rm dm}^2}{3 s^2} + \dfrac{4 m_f^2 + 4 m_{\rm dm}^2}{s} \right) \left( q_f^2 - \dfrac{g_V}{\cos^2 \theta_W} q_f \text{Re} \dfrac{s}{s - m_Z^2 - i m_Z \Gamma_Z} \right) \right. \\
        & + \dfrac{1}{4 \cos^4 \theta_W} \dfrac{1}{(s - m_Z^2)^2 + m_Z^2 \Gamma_Z^2} \left[ (g_V^2 - g_A^2) m_f^2 (4s^2 + 8m_{\rm dm}^2) \right. \\
        & + \left. \left. (g_V^2 + g_A^2)(s^2 + 4s m_{\rm dm}^2 - 8 m_{\rm dm}^2 m_f^2 + \dfrac{s^2 - 4 m_f^2 s - 4 m_{\rm dm}^2 s + 16 m_f^2 m_{\rm dm}^2}{3} \right] \right\}
    \end{split}
\end{align}
\normalsize
with $N_f^C$, $g_V$ and $g_A$ the color factor and Vector and Axial vector components of the SM fermions.

\subsection{Dark photon decay}
For the decay of dark photons into a pair of fermions,
\begin{equation}
    \Gamma'_{\gamma' \rightarrow f \overline{f}} = \dfrac{e^2 \cos^2 \theta_W Q_f^2 \varepsilon^2 N_c m'}{12 \pi}  \left( 1 + \dfrac{2 m_f^2}{m'^2} \right) \sqrt{1 - \dfrac{4 m_f^2}{m'^2}} \Theta(m' - 2 m_f)
\end{equation}
which we can sum over every particle that the dark photon is allowed to decay into (depending on its mass). For the decay into pairs of hadrons, we use the data extracted from ($e^+ e^- \rightarrow \rm hadrons$) \cite{Workman:2022ynf, Berger:2016vxi}:
\begin{equation}
    \Gamma'_{\gamma' \rightarrow \rm hadrons} = \Gamma'_{\gamma' \rightarrow \mu^+ \mu^-}  \times \mathcal{R} (m')
\end{equation}
where the ratio $\mathcal{R}$ is defined as
\begin{equation}
    \mathcal{R} \equiv \dfrac{\sigma(e^+ e^- \rightarrow \rm hadrons)}{\sigma(e^+ e^- \rightarrow \mu^+ \mu^-)}
\end{equation}
The total decay width of the dark photon is thereby
\begin{equation}
    \Gamma' = \sum_\ell \Gamma'_{\gamma' \rightarrow \ell \overline{\ell}} + \Gamma'_{\gamma' \rightarrow \rm hadrons}
\end{equation}
}
\section{Entropy production revisited}
\label{app:entropyProduction}
In the body of the text, we have used energy transfer from the HS to the VS to determine the evolution of $T'/T$ from which we could get the entropy produced. One can also write equations directly giving the evolution of the entropy, as in \cite{Scherrer:1984fd}. Both approaches are  equivalent, provided one takes into account the evolution of the effective degeneracy parameter of the VS. For simplicity,  we  assumed  $g_\ast$ constant during energy transfer for the analytical expressions but took into account its evolution for the numerical results. In this appendix, for the sake of comparison but also because of the specifics of our scenario, we derive some analytical expressions for entropy production, both for non-relativistic and relativistic  decay. We use lower case letters for the entropy densities and upper case for comoving ones. Primed (unprimed) quantities refer to the HS (resp. VS). Total quantities are written with a lower index $t$, e.g. $S_t = S' + S$ is the total comoving entropy, with $S_t = s_t a^3$. 

\subsection{Entropy production and non-relativistic reheating}
\label{sec:entropy_non_rel}

To set the ground, we start with the standard case of entropy production through the decay of a NR particle that comes to dominate the expansion of the Universe \cite{Scherrer:1984fd,Kolb:1990vq}. 
The heat transfer from decay satisfies 
\begin{equation}
 dQ = dE + pdV = - dQ'
\end{equation}
as the pressure of the NR decaying particles can be neglected. 
The comoving entropy of the VS  particles, all assumed to be relativistic, evolves as
\begin{equation}
{dS\over dt} \equiv  {1\over T} {d Q\over dt} = -{1\over T}{d(\rho' a^3)\over dt} 
\end{equation}
with $\rho' = m' n'$. Thus
\begin{equation}
{dS\over dt} = {\Gamma'\over T}{\rho' a^3} 
\end{equation}
with
\begin{equation}
\rho' a^3 = \rho'_i a^3_i e^{- \Gamma'(t -t_i)}
\end{equation}
Using MB statistics, $S = 4 g_\ast T^{3} a^3/\pi^2$ (appendix \ref{app:therm}), this can be written as
\begin{equation}
\label{eq:NRent}
S^{1/3}{dS\over dt} = \left({ 4 g_\ast\over\pi^2}\right)^{1/3}{ \Gamma'} \,\rho' a^4
\end{equation}
and
\begin{equation}
    \label{eq:entropyST}
S^{4/3} = S_i^{4/3} + {4\over 3} \left({ 4 \over\pi^2}\right)^{1/3}  \Gamma' \rho'_i a_i^4 \int_{t_i}^t dt' \, g_\ast^{1/3} {a\over a_i} e^{- \Gamma'(t' -t_i)}\nonumber
\end{equation}
Integrating this expression requires to know $a(t)$ and, possibly, the evolution of $g_\ast$ which depends on the temperature of the VS, $T$. We assume that the VS particles are always in equilibrium.  
If $g_\ast$ is constant, using $a\propto t^{2/3}$ for the early MD evolution, the entropy produced for $(t - t_i) \lesssim \tau' = 1/\Gamma'$ evolves as
\begin{equation}
\left({S\over S_i}\right)^{4/ 3} \approx 1+{3\over 5} {\rho'_i\over \rho_i} {\Gamma'\over H_i} \left(\left({t\over t_i}\right)^{5\over3} \!-\!1\right)
\end{equation}
where we have used MB statistics to express $s_i^{4/3}$ in terms of $\rho_i$.\footnote{Note that, for any statistics, $s^{4/3}\approx g^{1/3} \rho$ to within 1\%, regardless of the statistics (MB, FD or BE), see footnote \ref{foot:MB_approx}.} We see that $S \propto t^{5/4} \sim a^{15/8}$ for $a\propto t^{2/3}$ and thus $T \propto a^{-3/8}$ once the second term becomes dominant, see fig.\ref{fig:XiSchematic}. Notice that entropy production at early times depends on the heating parameter which we introduced in the bulk of this article, $\kappa \sim ({\rho'/\rho})(\Gamma'/H)$, see eq.\eqref{eq:heatpar}. 
At late times,  $t\gtrsim \tau'$, the entropy of the VS is
\begin{equation}
\left({S_f\over S_i}\right) \approx  (\Gamma({5/3}))^{3/4}  \left({\rho'_i \over \rho_i}\right)^{3/4} \left({\tau'\over t_i}\right)^{1/2}.
\label{eq:entropyST1}
\end{equation}
assuming $S_f \gg S_i$, with $\Gamma(5/3) \approx 0.9$. This is hardly new \cite{Scherrer:1984fd,Kolb:1990vq}. Nevertheless, a couple of features are  worth noticing for our problem. 

First, we notice that, as the ratio of matter to radiation evolves $\rho_{\rm m}/\rho_{\rm r} \propto a$ which is  $\sim t^{2/3}$ in a MD era, \eqref{eq:entropyST1} can be written  in terms of the ratio of energy densities at $\tau'$ as if the particle had not decayed  
\begin{equation}
{S_f\over S_i}\approx  \left({\rho'\over \rho}\right)^{3/4}_{\rm no decay}
\label{eq:entropyST2}
\end{equation}
Thus, the entropy produced is as if it had been stored in the decaying particle. 
Thus, the latter the decay, the larger is the entropy produced. The standard situation is the decay of massive particles that come to dominate the expansion at some $t_i = t_{\rm eq}$ \cite{Scherrer:1984fd,Kolb:1990vq}. At that moment, $\rho'_{\rm eq} = \rho_{\rm eq}$ and so the entropy produced is simply (and quite nicely) given by
\begin{equation}
{S_f\over S_{\rm eq}}\approx  \left({\tau'\over t_{\rm eq}}\right)^{1/2}
\label{eq:entropyST3a}
\end{equation}

In the bulk of this work, we consider a scenario in which a particle (a dark photon) becomes NR while it is driving the expansion of the Universe.  In that case, $t_i = t_{\rm nr}$ and the entropy of the VS is related to that in the HS by $S_{\rm nr} = (g_\ast/g'_\ast \xi^3)_{\rm nr}\, S'_{\rm nr} \ll S'_{\rm nr}$ where $\xi = T'/T \gg 1$. Thus, in that situation,  the total entropy produced is given by 
\begin{equation}
    {S_{t,f}\over S_{t,i}} \approx  {S_{f}\over S'_{i}} \approx \left({g_\ast\over g_\ast'}\right)^{1/4} \left({\tau'\over t_{\rm nr}}\right)^{1/2}
    \label{eq:entropyHHS}
\end{equation}

Alternatively, we can consider a scenario in which the particle is subdominant when it becomes NR but eventually dominates the Universe at $t_{\rm eq}$ before decaying. Between $t_{\rm nr}$ and $t_{\rm eq}$, $\rho \propto a^{-4} \propto t^{-2}$ and $\rho' \propto a^{-3} \propto t^{-3/2}$, so that $t_{\rm nr}/t_{\rm eq} = (\rho'_{\rm nr}/\rho_{\rm nr})^2$. From (\ref{eq:entropyST3a}), the entropy produced reads
\begin{equation}
    \frac{S_{t, f}}{S_{t, i}} \approx \frac{\rho'_{\rm nr}}{\rho_{\rm nr}} \left( \frac{\tau'}{t_{\rm nr}} \right)^{1/2}
    \label{eq:entropy_2}
\end{equation}

Finally, we can give the entropy produced if the massive particle is decaying when the expansion is RD, driven by the VS, a case which is relevant if $\xi  \ll 1$. 
Replacing $a \propto t^{2/3}$ by $a \propto t^{1/2}$ in \eqref{eq:entropyST} gives 
\begin{eqnarray}
\label{eq:entropyST1RD}
  \left({S_f\over S_i}\right)^{4/3}\!\!\! &\approx & 1 + \Gamma\!\left({3/2}\right)\! {\rho_i' \over \rho_i} \,\left({\tau'\over t_i}\right)^{1/ 2}
 \end{eqnarray}
 The second term can be expressed as the ratio of energy densities at decay assuming no entropy release, since $\rho_m/\rho_r \propto t^{1/2}$ in a RD era, and so is a small contribution as long as $\rho \gtrsim \rho'$.

\subsection{Entropy production and relativistic reheating}
Next we consider  entropy production when both HS and VS are made of relativistic particles, which requires to take into account the work done by the  fluids as the volume changes. The heat transfer satisfies
\begin{equation}
dQ = -dQ' \equiv - (dE' + p'dV)
\end{equation}
with $p' = \rho'/3$ and so  the rate  of entropy increase of the VS is given by
\begin{equation}
{dS\over dt}  = -{1\over a T}{d(\rho' a^4)\over dt}
\end{equation}
From section \ref{sec:history}, we can write this as
\begin{equation}
{dS\over dt} ={m' \Gamma' \over 3T'  T} {\rho' a^3}
\label{eq:entropyRel1}
\end{equation}
neglecting inverse decay. 
Notice that the entropy of the HS evolves as
\begin{equation}
{dS'\over dt} = - {m' \Gamma' \over 3 T'^2} {\rho' a^3},
\end{equation}
with no factor of $T \ll T'$, so the total entropy increases: $\dot S_t  >0$. 

With  $s = 4 g_\ast T^{3}/\pi^2$ (MB!), we can rewrite \eqref{eq:entropyRel1} as
\begin{equation}
S^{1/3}{dS\over dt} = \left({ 4 g_\ast\over\pi^2}\right)^{1/3}{m'\Gamma' \over 3T'} \,\rho' a^4
\end{equation}
This is the same expression as \eqref{eq:NRent}, provided $\Gamma' \rightarrow {m'\Gamma'/ 3 T'}$. 
This  can be integrated analytically for all $t$, provided $g_\ast$ = const and  $\rho'\gg \rho$. First, with $T' \propto 1/a$, we have 
\begin{equation}
\rho'  = \rho_i' {a^4_i\over a^4} e^{- {4 m' \Gamma'\over 9 T'_i H_i}\left((a/a_i)^{3} -1\right)},
\end{equation} 
see Eq.\eqref{eq:DeltaEvol}. Next, using MB statistics to express $s_i$ in terms of $\rho_i$, we get
\begin{equation}
    \label{eq:entropyCKT}
\left({S\over S_i}\right)^{4/3} \approx 1 + \frac{1}{3}{\rho'_i\over \rho_i} \left(1 -e^{- {4 m' \Gamma'\over 9 T'_i H_i}\left((a/a_i)^{3/2} -1\right)}\right)
\end{equation}
At early times,  entropy production depends  on the heating parameter $\kappa_i = (\rho'_i/3 \rho_i) \langle \Gamma'_i\rangle/H_i$ introduced in eq. \eqref{eq:heatpar}, 
\begin{equation}
\left({S\over S_i}\right)^{4/3} \approx 1+ \frac{8}{9} \kappa_i \left({a\over a_i}\right)^3 
\end{equation}
taking $a \gg a_i$.
It becomes significant only after contact, $\kappa_i \left({a_c\over a_i}\right)^3 \approx 1$ (see \ref{sec:attractor}) and then grows slowly as $S \propto  t^{9/8}\sim  a^{9/4}$ as $a\propto t^{1/2}$, 
corresponding to a temperature of the VS that evolves as $T\propto a^{-1/4}$, see fig.\ref{fig:XiSchematic}. After thermalization, $\langle \Gamma'\rangle/H \gtrsim 1$, and 
\begin{equation} 
\left({S_f\over S_i}\right) \approx \left({\rho_i'\over \rho_i}\right)^{3/4} \rightarrow \quad{S_{t,f}\over S_{t,i}} \approx {S_f\over S_i'} \approx \left({g_\ast \over g_\ast'}\right)^{1/4}
\label{eq:entropyProdRel}
\end{equation} 
This means that the energy has been transferred from the HS to the VS, 
$
g_\ast T_f^4 a_f^4 \approx {g_\ast'} T'^4_i a_i^4
$ \cite{Coy:2021ann}. Unlike the case of decay of a non-relativistic particle, the entropy produced is independent of the decay rate, see eq.\eqref{eq:entropyHHS}. Matching with \eqref{eq:entropyProdRel} is obtained by setting $t_{\rm nr} \approx \tau'$, corresponding to a HS particle that would become non-relativistic right at the time of thermalization. The entropy production in the process of energy transfer between the two RD sectors is quite mild, $S_f/S'_i \approx 2.4 $ for  $g'_\ast = 3$ and $g_\ast \approx 100$. This is to be contrasted with the case of the decay of a NR particle, in which the entropy produced grows with the particle lifetime.  This is essentially due to the fact that $\rho'/\rho \propto a$ for NR particles in the HS  while it is constant for relativistic particles, see eq.\eqref{eq:entropyST2}.

\section{Boltzmann equations}
\label{app:DEs}
In this appendix, we address some aspects of the derivation of the  Boltzmann equations that govern the coupled HS-VS system, eqs.\eqref{eq:continuity}, \eqref{eq:HSnumber} and \eqref{eq:HSenergytransfer}. 

 The first equation, eq. \eqref{eq:continuity}, expresses total energy conservation and so has no collision term. It involves the total equation of state $w_{\rm t}$
\begin{equation}
    w_{\rm t} = \frac{p+p'}{\rho + \rho'} = w' \frac{\rho'}{\rho_{\rm t}} + \frac{\rho_{\rm t} - \rho'}{3 \rho_{\rm t}} \, ,
\end{equation}
using $w = 1/3$, since we are always concerned with times during which the VS is radiation-dominated. The HS equation of state (eos) $w'$  is discussed toward the end of this appendix. 

{The second  equation concerns the evolution of the HS number density, \eqref{eq:HSnumber}. It involves only decay and inverse decay, $\gamma' \leftrightarrow f \overline{f}$, but it must account for the different temperatures of the two sectors, $T$ and $T'$,
\begin{equation}
    \frac{dn'}{dt} + 3 H n' = \int d\Pi_1 d\Pi_2 d\Pi' \vert \mathcal{M} \vert^2 (2 \pi)^4 \delta^4(p_1 + p_2 - p') \times (f_1 f_2 - f') \, ,
    \label{eq:B1}
\end{equation}
where  $d\Pi = d^3p/(2\pi)^3 2E$ \cite{Kolb:1990vq}. As usual, we assume that the SM fermions are in equilibrium with MB distributions at temperature $T$, $f_1 f_2 = f_{1,\rm eq} f_{2, \rm eq}= e^{-(E_1 + E_2)/T} {= f'_{\rm eq}(T)}$. In the HS, we allow for a possible departure from thermal equilibrium with temperature $T'$ through a chemical potential for the dark photons, $f' = e^{(\mu'-E')/T'} = e^{\mu'/T'}f'_{\rm eq} =  f'_{\rm eq}\, n'/n'_{\rm eq}(T')$. Using this and integrating over the momenta of the outgoing SM particles, equation \eqref{eq:B1} becomes
\begin{equation}
    \dfrac{dn'}{dt} + 3Hn' = \int {d^3 p'\over (2 \pi)^3} {m'\over E'} \Gamma' \left(f'_{\rm eq}(T) - f'_{\rm eq} n'(T')/n'_{\rm eq}(T') \right)
\end{equation}
where $\Gamma'$ is the dark photon decay rate in its rest frame. Integrating over the dark photon distributions at $T$ and $T'$ finally gives eq.\eqref{eq:HSnumber}, 
\begin{equation}
\frac{dn'}{dt} + 3Hn' = \langle \Gamma' \rangle_T \, n_{\text{eq}}(T) - \langle \Gamma' \, \rangle_{T'} n'(T') 
\label{eq:nprimeagain}
\end{equation}
where the thermal averages take into account time dilation at temperatures $T^{(\prime)} \gtrsim m'$, 
\begin{equation}
    \langle \Gamma' \rangle_{T^{(\prime)}} =  
    \Gamma' \left\langle{m'\over E'} \right\rangle_{T^{(\prime)}} =\Gamma' \frac{K_1(m'/T^{(\prime)})}{K_2(m'/T^{(\prime)})} \, .
    \label{eq:thermal_average}
\end{equation}

The last equation, eq.\eqref{eq:HSenergytransfer}, concerns energy transfer between the VS and the HS, 
\eqref{eq:HSenergytransfer}, 
\begin{equation}
    \frac{d\rho'}{dt} + 3(1+w') H \rho' = \int d\Pi_1 d\Pi_2 d\Pi' \, \,\vert \mathcal{M} \vert^2 (2 \pi)^4 \times \delta^4(p_1 + p_2 - p') (f_1 f_2 - f') E'  \, ,
\label{eq:B2}
\end{equation}
see e.g. \cite{Chu:2011be}.
As above, the collision term is readily integrated using equilibrium distributions,  
\begin{equation}
    \frac{d\rho'}{dt} + 3(1+w') H \rho' = \int {d^3 p'\over (2 \pi)^3} m' \Gamma' (f'_{\rm eq}(T) - f'_{\rm eq} n'(T')/n'_{\rm eq}(T'))
\end{equation}
where the factor of $E'$ in eq. \eqref{eq:B2} simplifies with the one from the measure $d\Pi'$. Integrating over dark photons distributions gives\footnote{The rhs of equation \eqref{eq:HSenergytransfer} only involves decay rate $\Gamma'$. The factor of $m'$ is reasonable on dimensional grounds, but it may be worth noticing that it also properly takes into account the suppression of energy transfer at high temperatures. Indeed, neglecting the expansion rate, \eqref{eq:HSenergytransfer} gives that the relative rate of change of temperature is $\dot T'/T' \propto \Gamma' m'/T'$ using $\rho' \propto T'^4$  and $n' \propto T'^3$.}
\begin{equation}
    \frac{d\rho'}{dt} + 3 ( 1 + w') H \rho'= m' \Gamma' \left[ n'_{\rm eq}(T) - n' \right]
\end{equation}
}

Eqs.\eqref{eq:continuity}, \eqref{eq:HSnumber} and \eqref{eq:HSenergytransfer} give three coupled differential equations for $\rho$, $\rho'$ and $n'$. 
We solve them as functions of the scale factor, $a$, rather than of time, $t$, by using $d/dt = (aH)d/da$. 
When the dark photon is relativistic, one can change variables by using $T' da = -a dT'$, and similarly when it is non-relativistic we have $2T' da = - a dT'$. 
However, the relation between $a$ and $T'$ at intermediate times is more involved in our scenario, since there may also be entropy production, in particular if the dark photons become non-relativistic. 

{To solve the Boltzmann equations, we also need the evolution of eos $w$ and $w'$ in terms of the scale factor. 
In the visible sector,  $w = 1/3$ always since the VS is radiation dominated before BBN. The eos of the HS $w'$ is more involved, since  the HS consists of particles (the dark photons) that may become non-relativistic along the evolution of the Universe. The problem  is that we need to  track the change of $w'$ between the relativistic and non-relativistic regimes to solve the Boltzmann equations.  Now, using the equilibrium expressions for pressure and energy density, $w'$ can be written as 
\begin{equation}
    w' = {p'\over \rho'} \equiv {1\over 3} \left( 1 - {m' n'\over \rho'} \left\langle {m'\over E'}\right\rangle\right)
    \label{eq:eosExact}
\end{equation}   
with $\rho' = n' \langle E'\rangle$, a function that changes smoothly from $w' = 1/3$ to $w'=0$. To effectively evaluate this quantity, we rely on Maxwell-Boltzmann statistics, as summarized in appendix \ref{app:therm}. The main advantage of using MB  statistics is that $w'$ does not depend on the chemical potential but only on $T'$. From there, one way to proceed is to solve for $T'$  (and thus for the pressure $p'$, $\langle m/E'\rangle$, etc.) at each step from given values of $n'$ and $\rho'$. A more efficient way is to use an approximate expression for $w'$ which depends only on $\rho'$ and $n'$, in which case  the system of Boltzmann equations \eqref{eq:continuity}, \eqref{eq:HSnumber} and \eqref{eq:HSenergytransfer} becomes closed. We use 
\begin{equation}
   w' \simeq \frac{1}{3} \left[ 1 - \left( \frac{m' n'}{\rho'} \right)^2 \right]\, ,
    \label{eq:approx_w}
\end{equation}
which stems from \eqref{eq:eosExact} by $\langle m'/E'\rangle \rightarrow m'/\langle E'\rangle =  m'n'/\rho'$. The ratio $\langle m'/E'\rangle/(m'/\langle E'\rangle) = 2/3$ for $T' \gg m'$ (see appendix \ref{app:therm}) but in that case $w' \approx 1/3$; for $T' \ll m'$, it tends exponentially fast towards 1, giving $\omega' \approx 0$. Consequently, the approximate expression \eqref{eq:approx_w} interpolates smoothly between $w' = 1/3$ and $w' = 0$ and differs from \eqref{eq:eosExact} by at most $5 \%$. 
Using this  approximation for $w'$, the parameter $x' \equiv m'/T'$ which enters explicitly in eq.\eqref{eq:HSnumber} is given by
\begin{equation}
    x' = {m'\over T'} = \frac{m'n'}{p'} = \frac{m'n'}{w' \rho'}
\end{equation}
We have checked  that the numerical solutions obtained using the iterative approach alluded to above and the approximation \eqref{eq:approx_w} give results that are in excellent agreement. The results we report are those obtained  using eq.\eqref{eq:approx_w}.
Further comments on the validity of these and related approximations that we use in our work are given in appendix \ref{sec:numsols}.}

\section{Comments on numerical solutions}
\label{sec:numsols}

In the body of the text, we focused on approximate analytical solutions and their comparison to the numerical solutions  depicted in fig.\ref{fig:evolution}. Here we add some brief comments on our numerical resolution of eqs.(\ref{eq:continuity}-\ref{eq:HSenergytransfer}).
 We solve for $\rho'$, $n'$ and $\rho$ and define from these quantities the temperature $T$ of the VS,  a proxy for the same quantity $T'$ for the HS and finally the chemical potential $\mu'$. Other quantities, like entropy densities, are obtained using standard but general equilibrium relations, eg $s' = (\rho' + p' - \mu' n')/T'$. 
For $T' \hat = \langle E'\rangle/3$ (MB),  we found convenient to define it through $T' = p'/n' = w' \rho'/n'$, where the HS equation of state, $p' = w' \rho'$, is approximated by (\ref{eq:approx_w}), see appendix \ref{app:DEs}.  As the system evolves, the dark photons can develop a non-zero chemical potential $\mu'$, which can be read out from $n'$ or $\rho'$ once $T'$ is determined. This is the case when the dark photons become non-relativistic as they are free streaming, and so, more abundant than their thermal equilibrium value. In that case, the temperature evolves as $T' \approx 1/a^2$ and their chemical potential as $\mu' - m' \propto T'/T'_{\rm nr}$ \cite{Kolb:1990vq}. 

A non-zero chemical potential can also arise for relativistic dark photons. This is in particular the case when the HS and the VS approach thermalization. This is related to the bumpy features around $\xi$ approaches $1$ that are visible in the numerical solutions (see the left panels of fig.~\ref{fig:evolution}). To gain some understanding of  the origin of these features, we combine  Eqs.\eqref{eq:HSnumber} and \eqref{eq:HSenergytransfer}. 
Neglecting the inverse process and using $K_1(x')/K_2(x') \approx x'/2$  and  $\rho'/n' \approx  3T'$ 
gives the following equation for $T'$, 
\begin{equation}
    \frac{dT'}{da} + \frac{T'}{a} \approx \frac{\Gamma'm'}{6 Ha}  
    \label{eq:rhonDP}
\end{equation}
Notice that the rhs is positive. 
As long as $\langle \Gamma'\rangle/H\ll 1$, this term can be neglected and the temperature evolves as $T' \propto a^{-1}$, corresponding to the attractor evolution, see section \ref{sec:attractor}.
Eventually, the rhs becomes significant, $\langle \Gamma'\rangle/H\sim 1$, and so $T'$ decreases more slowly than $a^{-1}$, leading to a levering of $\xi = T'/T$ close to thermalization and so to the bumpy feature  visible for instance in the lower left panel of fig.\ref{fig:evolution}. This is due to time dilation, which at late time leads to a depletion of the $\gamma'$ particle distribution at low energies and so, effectively, to an increase of their mean energy $\sim T'$.  This transient period does not last long, as $T'/T$ approaches 1 exponentially fast at the time of thermalization, introducing a slight delay in the thermalization process if $g_\ast \gg g_\ast'$, see section \ref{sec:rel_eq}. A more detailed but preliminary analysis, based of the evolution of the $\gamma'$ particle distribution, is in progress \cite{HufnagelKT}; it basically supports the validity of the above assertions.

\section{On freeze-in and thermalization of dark photons}
\label{app:FI}

In this appendix we consider the case $\xi_i \ll 1$. So we assume that the expansion is radiation dominated and driven by the VS, $T \gg T'$, and consider freeze-in production of dark photons and their subsequent decay. We do so to assess the possible impact of FI on the initial temperature ratio, $\xi_i \ll 1$. 
If $T \gg T'$, it is in principle required to take into account modifications of the ordinary photon propagation modes due to thermal corrections to accurately track the freeze-in production of dark photons \cite{An:2013yfc,Dvorkin:2019zdi,Hambye:2019dwd}. Nevertheless, as the final abundance of dark photons  is  dominated by the inverse decay process \cite{Berger:2016vxi,Hambye:2019dwd}, to simplify our discussion, in this brief appendix we neglect these subtleties and just take into account inverse decay to estimate the late abundance of dark photons.\footnote{If $T' \gg T$, we deem that thermal effects are negligible in the process of reheating of the VS. First, the dark photons are essentially non-interacting. Second the VS is much colder than the HS so that the impact of thermal corrections, such as modification of the mass of the SM degrees of freedom is a small disturbance to the process of reheating of the VS.}

The expansion is RD and driven by the VS if $\xi_i \ll 1$ but also more generally when $\rho' \ll \rho$ along the evolution of the coupled VS and HS. In particular, this is the case both after relativistic thermalization  or when the dark photons are non-relativistic but already decaying, $t \gtrsim \tau'$, so that their abundance is exponentially suppressed. In all these cases, we may rewrite the equation \eqref{eq:HSnumber} for the abundance of DP  as
\begin{eqnarray}
\frac{dY'}{dt} + \sigma' x Y' &\approx&  {g'\over 8 g_\ast} \sigma'  x^3 K_1(x)
\end{eqnarray}
where $x= m'/T$, $Y' = n'/s$ and $\sigma' = \Gamma'/H'  \lesssim 1$ with $H' = H(m')$. The lhs takes into account dark photon decay while the rhs is the source term from FI production. As before in the present work, we have used MB statistics to express $n'_{\rm eq}$  and $\langle \Gamma'\rangle$  in terms of Bessel functions, cf appendix \ref{app:therm}.  This Boltzmann equation is readily integrated to give
\begin{eqnarray}
    Y' &=& Y'_i \,e^{- {\sigma'\over 2} (x^2-x_i^2)} \nonumber\\
    &+& {g'\over 8 g_\ast} \sigma' \int_{x_i}^x dx'  x'^3 K_1(x') e^{-{\sigma'\over 2} (x^2-x'^2)}
    \label{eq:sol_FIlate}
\end{eqnarray}
The first term of the rhs represents the decay of the initial dark photon abundance, $e^{-\sigma' x^2/2} \sim e^{-t/\tau'}$; the second one represents their production from the VS convoluted by the dark photon decay. 

The simple expression \eqref{eq:sol_FIlate} captures several features that we observed in the more complicated problem of reheating of the VS from a HS. For the sake of brevity, we just sketch the key results, which can be readily verified by inspection of the general solution. For this, we refer to figure \ref{fig:FI_ex} in which some typical solutions are depicted. 

Consider first the solid lines, say for $\xi_i = 10^{-3}$. The abundance is initially characterized by a plateau $Y' \approx Y_i' \propto \xi_i^3$, analogous to the plateau in the case of a hot HS. For such choice of parameters, the initial temperature ratio is stable and we can discuss the freeze-out of DM along the line of section \ref{sec:domain}, with $\xi_i = \xi_{\rm fo}$. Eventually, freeze-in production of dark photons becomes relevant. This occurs essentially when the heating parameter $(1/Y') \langle\Gamma'\rangle/H$ becomes $\mathcal{O}(1)$, marking the onset of dark photon creation from the VS, after which $Y' \propto a^3$ (and so $\xi \propto a$). This combination of parameters is akin to the heating parameter $\kappa \sim (\rho'/\rho) \langle \Gamma'\rangle/H$ in the problem of heating of the VS from the HS.\footnote{There is a slight difference between the criteria for the onset of particle creation $\sim (1/Y') \langle \Gamma'\rangle/H\sim 1$ and that of  energy transfer $\kappa'\sim \rho/\rho'\langle \Gamma'\rangle/H \sim 1$ (heating parameter). For freeze-in, we deem more relevant to focus on the number density of dark photons rather than on their mean energy. For the problem of reheating of the VS, the key issue is of course that of energy transfer.} Different choices of $\xi_i$ depict the same behaviour and for all they track the production of dark photons curve, which in that respect behaves as an attractor. Such behavior includes cases in which $\xi_i$ is very low. For instance, for $\xi_i=  10^{-6}$, the abundance of dark photons produced by freeze-in increases rapidly (see solid purple line) toward this attractor solution. This sharp initial increase of the number of dark photons is analogous to the rapid increase of temperature to $T_{\rm max}$ in the problem of reheating of the VS. This occurs when $(1/Y_i) \langle\Gamma'\rangle_i/H_i' \gtrsim 1$ at the initial moment. In that case, the initial choice of $\xi_i$ is unstable. In the body of the text and in figure \ref{fig:newDomain} we require that 
\begin{equation}
    \kappa'_i = \frac{\rho_i}{3 \rho'_i} \frac{\langle \Gamma' \rangle_i}{H_i} \lesssim 1
\end{equation}
for the stability of the initial temperature ratio when $\xi_i\lesssim 1$, see section \ref{sec:stability}.

After freeze-in production, the dark photons are non-relativistic and $Y' \sim $ constant until decay becomes relevant. As the solid and dotted curves show, the largest the freeze-in production, the earlier the decay of the dark photons. That trivially results from the fact that their creation and disappearance are controlled by related processes (inverse and direct decay), in other words by the kinetic mixing parameter. In particular, a dark photon that would reach $Y'\sim Y'_{\rm eq}$ at $T \sim m$ would, by definition, be in thermal equilibrium and would subsequently track their equilibrium abundance, $n' \propto e^{-m'/T}$, see the dot-dashed curve in fig.\ref{fig:FI_ex}. For a smaller kinetic mixing, their abundance will overshoot the equilibrium abundance until they start to decay, $n' \propto e^{-\Gamma' t}$, see solid and dotted curves. As is known since some time, see \cite{Harvey:1981yk}, such dark photons may eventually thermalize with the VS if $n' \propto e^{-\Gamma't} \sim n'_{\rm eq}(T)$, see solid curves and their merging with the equilibrium abundance. Such outcome is of course only relevant if the number density $n'$ is not a ridiculously small number when that condition is met, see short dashed curves.
\begin{figure}[t!]
    \centering
    \includegraphics[width=0.99\columnwidth]{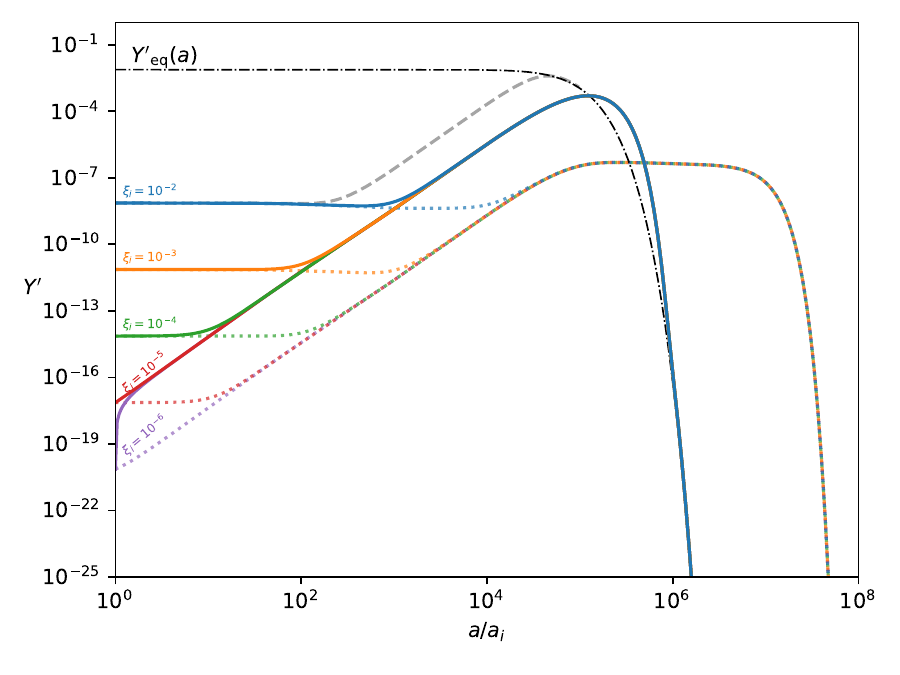}
    \caption{{Examples of evolution of dark photons abundance for $\xi_i \ll 1$. The mass $m'$ is fixed at 5 MeV. The solid curves correspond to a value for $\varepsilon$ of $10^{-9.4}$, only differing by the value of $\xi_i$, and the long-dashed one to an $\varepsilon = 10^{-8.5}$. The dotted curves are obtained by setting $\varepsilon$ to $10^{-11}$. One can observe the attractor behavior for the curves that share the same $\varepsilon$ and, afterwards, the contact point between $Y'$ and its equilibrium value $Y'_{\rm eq}$ for this choice of parameters. }}
    \label{fig:FI_ex}
\end{figure}

\newpage

\bibliographystyle{apsrev}
\bibliography{peakbib}


\end{document}